\documentclass[twocolumn]{aastex63}
\usepackage{graphicx}
\usepackage{float}
\usepackage{amsmath,amssymb}
\usepackage{xcolor}
\usepackage{multirow} 

\def\Msol{\rm{M_{\odot}}}

\def\sn{AT\,2020xnd}

\usepackage{colortbl}

\received{}
\revised{}
\accepted{}
\submitjournal{ApJ}

\shorttitle{AT\,2020xnd}
\shortauthors{Bright et al.}
\graphicspath{{./}{figures/}}

\begin{document}

\title{Radio and X-ray observations of the luminous Fast Blue Optical Transient AT\,2020xnd}

\correspondingauthor{Joe S. Bright}
\email{joe.bright@berkeley.edu}

\author[0000-0002-7735-5796]{Joe S. Bright}
\affiliation{Department of Astronomy, University of California, Berkeley, CA 94720-3411, USA}
\affiliation{Center for Interdisciplinary Exploration and Research in Astrophysics (CIERA) and Department of Physics and Astronomy, Northwestern University, Evanston, IL 60208}

\author[0000-0003-4768-7586]{Raffaella Margutti}
\affiliation{Department of Astronomy, University of California, Berkeley, CA 94720-3411, USA}
\affiliation{Center for Interdisciplinary Exploration and Research in Astrophysics (CIERA) and Department of Physics and Astronomy, Northwestern University, Evanston, IL 60208}

\author{David Matthews}
\affiliation{Department of Astronomy, University of California, Berkeley, CA 94720-3411, USA}
\affiliation{Center for Interdisciplinary Exploration and Research in Astrophysics (CIERA) and Department of Physics and Astronomy, Northwestern University, Evanston, IL 60208}

\author{Daniel Brethauer}
\affiliation{Department of Astronomy, University of California, Berkeley, CA 94720-3411, USA}
\affiliation{Center for Interdisciplinary Exploration and Research in Astrophysics (CIERA) and Department of Physics and Astronomy, Northwestern University, Evanston, IL 60208}

\author[0000-0001-5126-6237]{Deanne Coppejans}
\affiliation{Center for Interdisciplinary Exploration and Research in Astrophysics (CIERA) and Department of Physics and Astronomy, Northwestern University, Evanston, IL 60208}
\affiliation{Department of Physics, University of Warwick, Coventry CV4 7AL, UK}

\author[0000-0002-7721-8660]{Mark H. Wieringa}
\affiliation{CSIRO Space \& Astronomy, PO Box 76, Epping NSW 1710, Australia}

\author[0000-0002-4670-7509]{Brian D. Metzger}
\affil{Department of Physics and Columbia Astrophysics Laboratory, Columbia University, Pupin Hall, New York, NY 10027, USA}
\affil{Center for Computational Astrophysics, Flatiron Institute, 162 5th Ave, New York, NY 10010, USA} 

\author[0000-0003-4587-2366]{Lindsay DeMarchi}
\affiliation{Center for Interdisciplinary Exploration and Research in Astrophysics (CIERA) and Department of Physics and Astronomy, Northwestern University, Evanston, IL 60208}

\author[0000-0003-1792-2338]{Tanmoy Laskar}
\affiliation{Department of Physics, University of Bath, Claverton Down, Bath, BA2 7AY, UK}
\affiliation{Department of Astrophysics/IMAPP, Radboud University Nĳmegen, P.O. Box 9010, 6500 GL Nĳmegen, The Netherlands}

\author[0000-0001-5725-0359]{Charles Romero}
\affiliation{Center for Astrophysics | Harvard \& Smithsonian, Cambridge, MA 02138, USA}
\affiliation{Green Bank Observatory, P.O. Box 2, Green Bank, WV 24944}
\affiliation{Department of Physics and Astronomy, University of Pennsylvania, 209 South 33rd Street, Philadelphia, PA, 19104, USA}

\author{Kate D. Alexander}
\affiliation{Center for Interdisciplinary Exploration and Research in Astrophysics (CIERA) and Department of Physics and Astronomy, Northwestern University, Evanston, IL 60208}

\author{Assaf Horesh}
\affiliation{Racah Institute of Physics, The Hebrew University of Jerusalem, Jerusalem 91904, Israel}

\author[0000-0003-0216-8053]{Giulia Migliori}
\affiliation{Istituto di Radioastronomia - INAF, Via P. Gobetti 101, I-40129 Bologna, Italy}

\author[0000-0002-7706-5668]{Ryan Chornock}
\affiliation{Department of Astronomy, University of California, Berkeley, CA 94720-3411, USA}


\author[0000-0002-9392-9681]{E.~Berger}
\affiliation{Center for Astrophysics | Harvard \& Smithsonian, Cambridge, MA 02138, USA}

\author{Michael Bietenholz}
\affiliation{Department of Physics and Astronomy, York University, Toronto, M3J 1P3, Ontario, Canada}
\affiliation{SARAO/Hartebeesthoek Radio Observatory, PO Box 443, Krugersdorp, 1740, South Africa}

\author{Mark J. Devlin}
\affiliation{Department of Physics and Astronomy, University of Pennsylvania, 209 South 33rd Street, Philadelphia, PA, 19104, USA}

\author[0000-0002-1940-4289]{Simon R.\ Dicker}
\affiliation{Department of Physics and Astronomy, University of Pennsylvania, 209 South 33rd Street, Philadelphia, PA, 19104, USA}

\author[0000-0002-3934-2644]{W.~V.~Jacobson-Gal\'{a}n}
\affiliation{Department of Astronomy, University of California, Berkeley, CA 94720-3411, USA}
\affiliation{Center for Interdisciplinary Exploration and Research in Astrophysics (CIERA) and Department of Physics and Astronomy, Northwestern University, Evanston, IL 60208}

\author{Brian S. Mason}
\affiliation{National Radio Astronomy Observatory, 520 Edgemont Rd., Charlottesville VA 22903, USA}

\author[0000-0002-0763-3885]{Dan Milisavljevic}
\affiliation{Department of Physics and Astronomy, Purdue University, 525 Northwestern Avenue, West Lafayette, IN 47907, USA}

\author[0000-0002-6154-5843]{Sara E. Motta}
\affiliation{Istituto Nazionale di Astrofisica, Osservatorio Astronomico di Brera, via E.\,Bianchi 46, 23807 Merate (LC), Italy}
\affiliation{University of Oxford, Department of Physics, Astrophysics, Denys Wilkinson Building, Keble Road, OX1 3RH, Oxford, United Kingdom}

\author[0000-0003-3816-5372]{Tony Mroczkowski}
\affiliation{ESO - European Southern Observatory, Karl-Schwarzschild-Str.\ 2, D-85748 Garching b.\ M\"unchen, Germany}

\author{Enrico Ramirez-Ruiz}
\affiliation{Department of Astronomy and Astrophysics, University of California, Santa Cruz, CA 95064}
\affiliation{Niels Bohr Institute, University of Copenhagen, Blegdamsvej 17, 2100 Copenhagen, Denmark}

\author{Lauren Rhodes}
\affiliation{Astrophysics, University of Oxford, Denys Wilkinson Building, Keble Road, Oxford, OX1 3RH, UK}
\affiliation{Max Planck Institute f\"{u}r Radioastronomie, Auf dem H\"{u}gel, Bonn 53121, Germany}

\author{Craig L. Sarazin}
\affiliation{Department of Astronomy, University of Virginia, 530 McCormick Road, Charlottesville, VA 22904-4325, USA}

\author{Itai Sfaradi}
\affiliation{Racah Institute of Physics, The Hebrew University of Jerusalem, Jerusalem 91904, Israel}

\author{Jonathan Sievers}
\affiliation{Department of Physics, McGill University, 3600 University Street Montreal, QC H3A 2T8, Canada}




\begin{abstract}
We present deep X-ray and radio observations of the Fast Blue Optical Transient (FBOT) AT\,2020xnd/ZTF20acigmel at $z=0.2433$ from $13$~d to $269$~d after explosion. AT\,2020xnd belongs to the category of optically luminous FBOTs with similarities to the archetypal event AT\,2018cow. AT\,2020xnd shows luminous radio emission reaching $L_{\nu}\approx8\times10^{29}$~erg\,s$^{-1}$Hz$^{-1}$ at 20GHz and $75$d post explosion, accompanied by luminous and rapidly fading soft X-ray emission peaking at $L_{X}\approx6\times10^{42}$~erg\,s$^{-1}$. Interpreting the radio emission in the context of synchrotron radiation from the explosion's shock interaction with the environment we find that AT\,2020xnd launched a high-velocity outflow ($v\sim$0.1--0.2$c$) propagating into a dense circumstellar medium (effective $\dot M\approx10^{-3}\Msol$~yr$^{-1}$ for an assumed wind velocity of $v_w=1000\,$km\,s$^{-1}$).  Similar to AT\,2018cow, the detected X-ray emission is in excess compared to the extrapolated synchrotron spectrum and constitutes a different emission component, possibly powered by accretion onto a newly formed black hole or neutron star. These properties make \sn{} a high-redshift analog to AT\,2018cow, and establish AT\,2020xnd as the fourth member of the class of optically-luminous FBOTs with luminous multi-wavelength counterparts.
\end{abstract}



\section{Introduction} \label{sec:intro}
The advent of wide-field and high-cadence optical transient surveys, along with real-time discovery efforts, has expanded the parameter space in the search for new classes of extragalactic transient with rapid evolution timescales. Observations from such surveys have revealed a variety of optical transients spending $\lesssim10$ days above half maximum brightness, atypical for the majority of extragalactic transients previously discovered (e.g., \citealt{Kasliwal10, Poznanski10}). Among these are the Fast Blue Optical Transients (FBOTs), characterized by their rapid rise to maximum light ($\lesssim10\,\rm{d}$), peak optical luminosity reaching $L_{pk}>10^{43}\,\rm{erg}\,s^{-1}$, and persistently blue colors (e.g. \citealt{Drout14,Arcavi16,Tanaka16,Pursiainen18,Rest18,ho2021,perley2020b}). FBOTs appear to be intrinsically rare events, occurring at between 1\% and 10\% of the core collapse supernova rate in the local Universe
\citep{Drout14, Pursiainen18, Tampo20, Li11_LOSS}.

\begin{figure*} 
	\centering
	\includegraphics[width=0.48\textwidth]{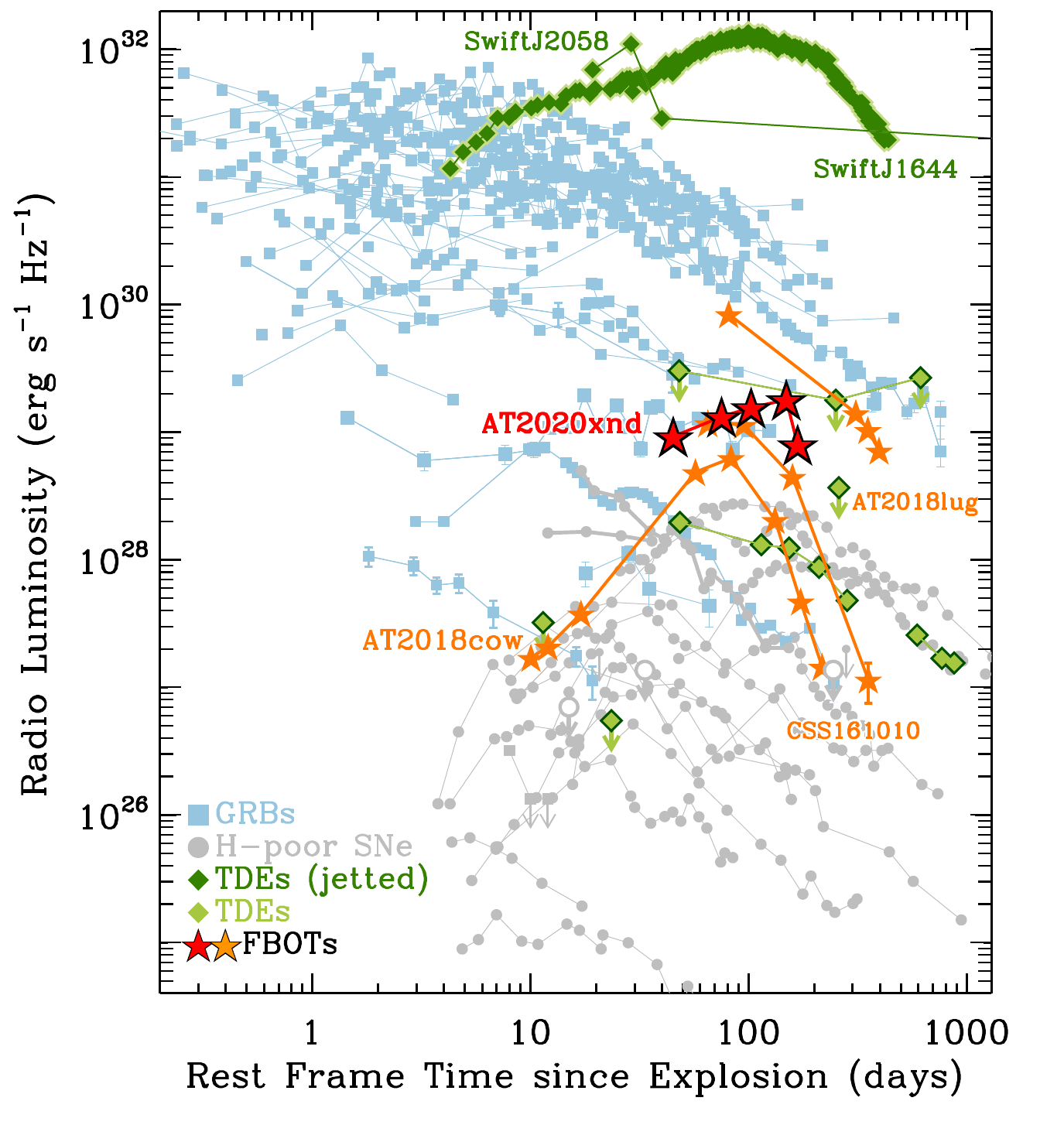}
	\includegraphics[width=0.48\textwidth]{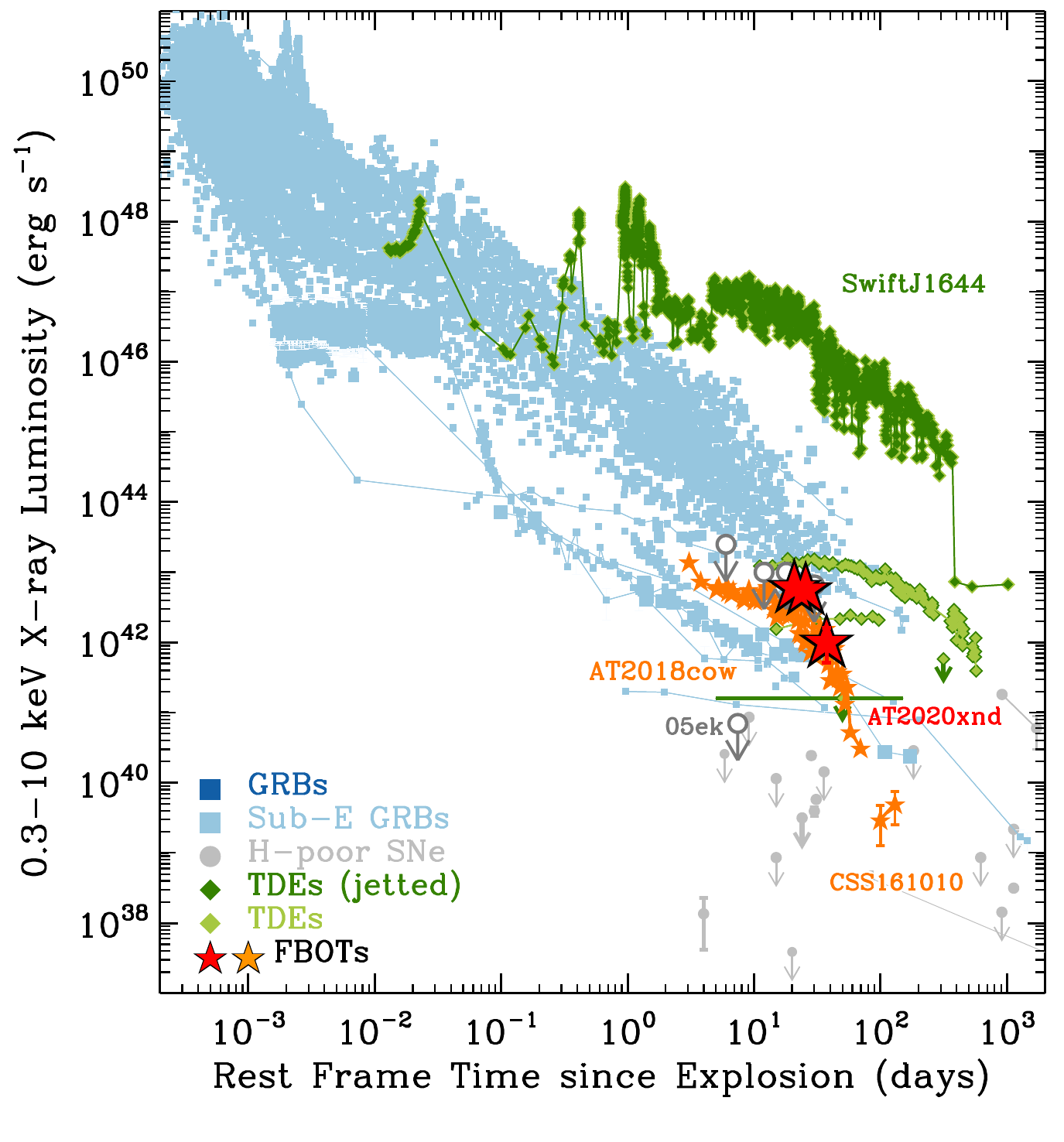}
    \caption{\emph{Left Panel:} The evolving radio luminosities of extragalactic H-poor explosive transients, including long GRBs (light blue squares), SNe (light-gray circles), TDEs (green diamonds)  at $\nu\approx 5$ GHz. Like the other FBOTs (orange stars), \sn{} (red stars) drops off rapidly at late times. \emph{Right Panel:} X-ray parameter space for the same set of transients.  References: \cite{Margutti19,Alexander20,Coppejans20,ho2020} and references therein.}
    \label{Fig:XrayFBOTS_population}
\end{figure*}

The most optically luminous FBOTs are further distinguished from other rapidly evolving extragalactic transients - such as subluminous Type IIb/Ib SNe \citep{Poznanski10,ho2021}; luminous Type Ibn or hybrid IIn/Ibn SNe \citep{ho2021}; Type IIn SNe (e.g. \citealt{Ofek10}) - based on the presence of highly luminous X-ray and radio emission, comparable to those seen in short gamma-ray bursts and well in excess of what is seen in typical core-collapse SNe (e.g. \citealt{Margutti19,Coppejans20,ho19b,ho_xnd_atel1,bright_xnd_atel1,matthews_xnd_atel1}, and see Figure \ref{Fig:XrayFBOTS_population}). Although the population of FBOTs with associated high energy emission remains small, FBOTs with luminous radio and X-ray emission are rarer still, occurring at $\lesssim1\%$ of the core-collape supernovae (CCSNe) rate below $z\sim0.5$ \citep{Coppejans20, ho2021}. In the optical, FBOTs do not show a $^{56}\rm{Ni}$ powered decay tail, do show high temperature photospheric emission, and are preferentially located in dwarf galaxies \citep{Coppejans20,Perley2021}. 

The prototypical optically luminous FBOT with associated emission at X-ray and radio wavelengths is AT2018cow (\citealt{Prentice18}) which, at only $\sim60\,\rm{Mpc}$, was the subject of extended observing campaigns at cm, mm, and X-ray wavelengths \citep{Margutti19,ho2018cow,Kuin19,Perley19,nayana2021,RiveraSandoval18}. The radio observations revealed the presence of a shock with velocity of $\sim0.1c$ interacting with a dense and asymmetric CSM, while the X-ray emission was in excess of an extrapolation of the radio spectrum. This X-ray excess suggested an additional emission component, which was interpreted as a central engine - an accreting compact object or a spin-powered magnetar. The X-ray spectrum of AT2018cow also clearly contained multiple components, with an excess above $\sim10\,\rm{keV}$ seen at early times that had vanished at $\sim15\,\rm{d}$ post explosion. The origin of this hard excess remains unclear, \citet{Margutti19} interpreted it as a Compton hump feature resulting from X-rays interacting with a fast ejecta shell, or reflection off of an accretion funnel. Both scenarios suggest an X-ray source embedded within the explosion. While X-ray observations of CSS161010 \citep{Coppejans20} at $d\approx 150$ Mpc were not as comprehensive as for AT2018cow, the former showed a similar excess relative to its well sampled radio SEDs, suggesting that the presence of a central engine is a feature of the FBOTs \citep{Margutti19,Coppejans20}. 

In this work we present radio and X-ray observations of the FBOT \sn{} (ZTF20acigmel), the third FBOT with both luminous X-ray and radio emission. Additional radio observations of AT2020xnd/ZTF20acigmel (including high frequency monitoring with the SMA/NOEMA) were obtained by an independent observing team and are presented in Ho et al.+2021. \sn{} was discovered on 2020 October 12 by the Zwicky Transient Facility (ZTF, \citealt{bellm2019,graham2019}) as part of the two-day cadence public survey (\citealt{Perley2021}). Follow-up observations identified a candidate (dwarf) host galaxy at $z=0.2433$. Based on the rapid rise and decay, the blue color, the dwarf galaxy host, and the high optical luminosity of this source \cite{Perley2021} classified \sn{} as an FBOT.

Upon the announcement that \sn{} was producing luminous radio emission \citep{ho_xnd_atel1}, we initiated a multi-wavelength observing campaign on the target beginning at fifteen days post discovery. Our campaign included observations at radio (AMI-LA, ATCA, eMERLIN, GMRT, MeerKAT, VLA), sub-mm/mm (ALMA, GBT), and X-ray (Chandra, XMM-Newton) frequencies. We particularly highlight the use of the MUSTANG-2 bolometer camera on the GBT, which provided us with early time mm data, demonstrating its suitability for rapid transient follow-up at high frequencies.

We structure the rest of this manuscript as follows. In Section 2 we describe our observations and the data reduction process. In Section 3 we derive and present the results for our analysis of our radio and X-ray observations. Finally, in Section 4 and Section 5 we discuss our results and give our conclusions, respectively.  

\section{Observations} \label{sec:observations}
Throughout this paper, measurements in time are in reference to the explosion date ($T_0$), which is MJD $59132.0$ \citep{Perley2021}, and are in the observed frame, unless otherwise specified.  Uncertainties are reported at the $1\sigma$ (Gaussian equivalent) confidence level (c.l.) and upper limits at the $3\,\sigma$ c.l. unless explicitly noted. We adopt standard $\Lambda$CDM cosmology with H$_0=69.6$\,km~s$^{-1}$~Mpc$^{-1}$, $\Omega_{\rm{M}}=0.286$, $\Omega_{\rm{\Lambda}}=0.714$ \citep{Bennett2014}. At $z=0.2433$ \citep{Perley2021} the luminosity distance of \sn{} is $D_{L}=1232\,\rm{Mpc}$ and the angular diameter distance is $D_{\theta}=797\,\rm{Mpc}$.

\subsection{Radio}
In this section we describe our large radio campaign on \sn{}. A summary of all of our radio observations are given in Table \ref{Tab:radio}. The calibrators used and array configurations are given in Table \ref{Tab:radio_cal}.

\subsubsection{Australia Telescope Compact Array} \label{subsec:atca}
The field of \sn{} was observed with the Australia Telescope Compact Array (ATCA) under project codes CX471 (PI Bright) and CX472 (PI Ho), beginning on 2020 October 25 ($T_0+15$d). 
Data were recorded with either the $4\,\rm{cm}$ receiver (which collects data simultaneously at 5.5 and $9\,\rm{GHz}$), the $15\,\rm{mm}$ receiver (which collects data simultaneously at 17 and $19\,\rm{GHz}$), or the $7\,\rm{mm}$ receiver (which collects data simultaneously at 33 and $35\,\rm{GHz}$). Each frequency was observed with a $2\,\rm{GHz}$ bandwidth and processed by the Compact Array Broadband Backend (CABB; \citealt{wilson2011}). Data taken as part of CX471 were reduced and imaged in \textsc{miriad} \citep{sault1995} using standard techniques. Data from CX472 are reported in \citealt{dobie_atel1, dobie_atel2, dobie_atel3}. For observations taken with the $15\,\rm{mm}$ or $7\,\rm{mm}$ receiver the sub-bands were jointly imaged in order to double the bandwidth and increase image sensitivity.

\subsubsection{Karl G. Jansky Very Large Array} \label{subsubsec:vla}
We initiated Karl G. Jansky Very Large Array (VLA) observations of \sn{} as part of program VLA/20A-354 (PI Margutti) beginning on 2020 November 5 (MJD 59158, $T_0+26$d). Observations were taken at S, C, X, Ku, K, and Ka bands, utilizing the WIDAR correlator, with a $2\,\rm{GHz}$ bandwidth at S-band, a $4\,\rm{GHz}$ bandwidth at C and X-band, a $6\,\rm{GHz}$ bandwidth at Ku-band, and a $8\,\rm{GHz}$ bandwidth at K-band and Ka-band. \sn{} lies in the declination range of the Clarke satellite belt as observed from the VLA, and as such we shifted the basebands at C and X bands to reduce the impact of radio frequency interference. Data were reduced with the VLA CASA (\citealt{McMullin07}) calibration pipeline version 2020.1.0.36 and then manually inspected, further flagged, and reprocessed through the pipeline. The final imaging was performed using WSClean \citep{offringa2014,offringa-wsclean-2017} where we used \textsc{-fit-spectral-pol=2} (equivalent to using two Taylor terms when using CASA) to account for the wide fractional bandwidth of the VLA. Images were created using a Briggs parameter between 0 and 1 depending on the array configuration. Where we measured the flux within small sub-bands of the bandwidth we set \textsc{-no-mf-weighting} in order to avoid the creation of artificial spectral structure. Fitting was performed using \textsc{pybdsf} \citep{mohan2015} with the size of the source fixed to that of the synthesized beam. 

\subsubsection{Enhanced Multi-Element Radio Linked Interferometer Network} \label{subsubsec:eMERLIN}
We were awarded DDT observations (project ID DD10005, PI Bright) of \sn{} with the Enhanced Multi-Element Radio Linked Interferometer Network (eMERLIN) and observed the field of \sn{} on 2020 November 6 ($T_0+27$d). Observations were conducted at a central frequency of $5.075\,\rm{GHz}$ with a $512\,\rm{GHz}$ bandwidth. The Lovell telescope was not included in the array. Data calibration was performed with the eMERLIN CASA pipeline using standard techniques. We did not detect emission consistent with the position of \sn{}. We triggered a further four C-band observations as part of project ID CY11008 (PI Bright). Further observations were reduced using the same strategy, with imaging performed manually on the pipeline output, and we detected the source in one epoch at $T_0+102$d.

\subsubsection{Robert C. Byrd Green Bank Telescope} \label{subsubsec:gbt}
We observed \sn{} with the Robert C. Byrd Green Bank Telescope (GBT) with the MUSTANG-2 instrument \citep{dicker2014} beginning on $T_{0}+28\,\rm{d}$ (MJD 59160). Data were taken under projects GBT20B\_437 (PI Bright) and GBT20B\_440 (PI Bright). MUSTANG-2 is a bolometer camera providing $9^{\prime\prime}$ arcsecond resolution and high continuum sensitivity between 75 and $105\,\rm{GHz}$. Due to the wide bandwidth that MUSTANG-2 is sensitive to, the effective central frequency of any given observation depends on the spectral index of the source being observed. The spectral index from \sn{} (as determined by our modeling in \S\ref{sec:results}) is between $\sim1.5$ and $0$ through the MUSTANG-2 bandpass, which results in a central frequency between 88.6 and $88.9\,\rm{GHz}$. We therefore use $90\,\rm{GHz}$ a the central frequency, and do not consider the small shift due to spectral index.
The MUSTANG-2 data is reduced via the MIDAS pipeline, which is described in \citet{romero2020}. The MIDAS pipeline relies on Fourier filtering the data and subtraction of principle components (via PCA). For point sources such as \sn{} the recovered flux density is insensitive to the range of typical filtering parameters used. We used six principal components and a notched Fourier filter, keeping frequencies between 0.07 and 41 Hz.
The MUSTANG-2 data were flux calibrated relative to Neptune and a nearby secondary calibrator (a point source) was used to track pointing and gain shifts during each night. The observations of the secondary calibrators were also used to determine the beam size for each night.

\subsubsection{Giant Metrewave Radio Telescope}
We observed the field of \sn{} with the Giant Metrewave Radio Telescope (GMRT) beginning on $T_{0}+162$d at $0.75$GHz and $1.25$GHz, under program 39\_034 (PI Matthews). The data were reduced manually using standard calibration techniques, with multiple rounds of phase-only and then amplitude and phase self-calibration performed. We then performed a single round of direction-dependent self-calibration using killMS \citep{tasse2014,smirnov2015} to solve for direction dependent gains and DDFacet \citep{tasse2018} to perform imaging and create a compatible sky model.

\subsubsection{MeerKAT}
We observed the field of \sn{} with the MeerKAT radio telescope as part of project SCI-20210212-JB-01 (PI Bright) starting at $T_0+191$d. Observations were taken at a central frequency of $1.28\,\rm{GHz}$ with a bandwidth of $856\,\rm{MHz}$ and the correlator in 4k mode (4096 spectral channels of width $209\,\rm{kHz}$). Observations were reduced using the \textsc{oxkat}\footnote{https://github.com/IanHeywood/oxkat} reduction pipeline (\citealt{heywood_oxkat,McMullin07,kenyon2018,kurtzer2017,offringa2014}) which is a set of semi-automated scripts that we use to perform phase-reference calibration and self-calibration.

\subsubsection{Atacama Large Millimeter Array}
Due to the COVID-19 pandemic we were not able to trigger Atacama Large Millimeter Array (ALMA) observations close to $T_{0}$, however we obtained late time observations of \sn{} under project code 2019.1.01157.T (PI Coppejans) at $T_0+$184d and $T_0+$269 d (Table \ref{Tab:radio}). The first epoch consisted of a band-3 and a band-4 observation, whereas the second epoch consisted of a single deep band-3 observation. We used the results of the ALMA pipeline (which uses standard techniques and the CASA package) to calibrate and image the data. \sn{} was not detected in any of our ALMA observations. 

\subsubsection{Arcminute Microkelvin Imager Large-Array}
We observed the field of \sn{} with the Arcminute Microkelvin Imager Large Array (AMI-LA; \citealt{zwart2008, hickish2017}) (PI Bright) beginning on $T_0+154$d. We observed the field on a further three occasions but did not detect emission from \sn{}. Data were reduced using standard techniques with the \textsc{reduce\_dc} software (e.g. \citealt{perrott2013,bright2018}).

\subsection{X-ray Observations}
\begin{figure} 
	\centering
	\includegraphics[width=0.48\textwidth]{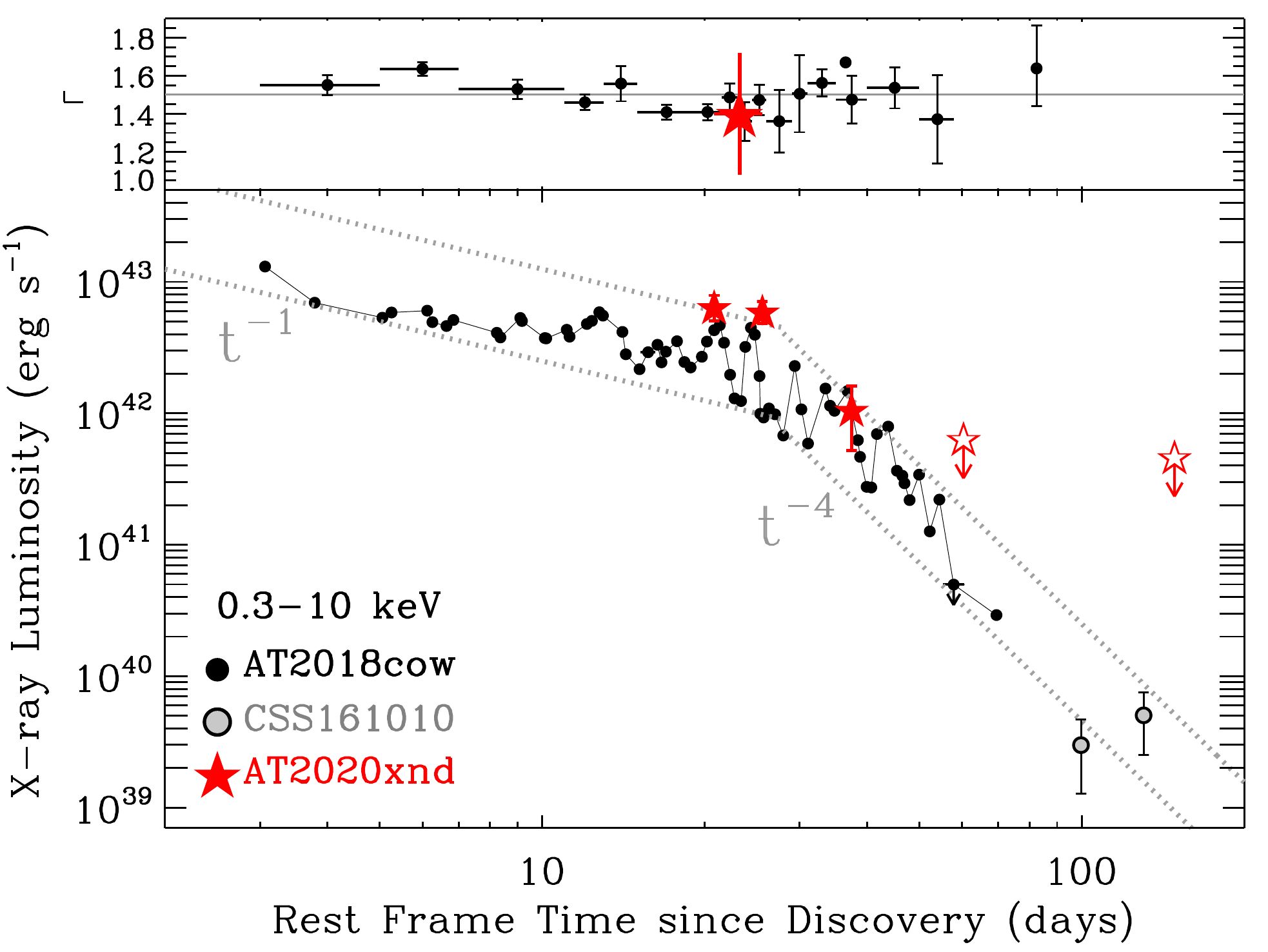}
    \caption{\emph{Main Panel:} Soft X-ray (0.3-10 keV) luminosity evolution of \sn{}  (red stars) compared to the other two known FBOTs with X-ray detections AT\,2018cow and  CSS\,161010 (filled black  and grey circles, respectively; \citealt{Margutti19,Coppejans20} ). 
    Dotted lines mark a $L_x\propto t^{-1}$ and a $L_x\propto t^{-4}$ power-law decay to guide the eye. Interestingly, \sn{} shows a similar luminosity as AT\,2018cow and seems to follow a similar, peculiar temporal evolution, with roughly constant X-ray flux at $\delta t\le 30$ d followed by a sharp decay. \emph{Upper panel}: spectral photon index evolution with time for AT\,2018cow and AT\,2020xnd. The solid horizontal line is a photon index of 1.5. 
    }
    \label{Fig:XrayFBOTS}
\end{figure}

The broad-band X-ray monitoring campaign described in this paper covered the time window $\delta t\sim$25--240$\,\rm{d}$ after explosion (Figure \ref{Fig:XrayFBOTS}).

\subsubsection{Chandra X-ray Observatory (0.3-10 keV)} \label{SubSec:Chandra}
We acquired deep X-ray observations of \sn{} with the \emph{Chandra} X-ray Observatory (CXO) 
under a joint CXO-NuSTAR program \#22500192 (PI Matthews; IDs 23547, 23548, 23549; 
exposure time of $\approx$19.8\,ks per ID) covering the time period $\delta t=25-240\,\rm{d}$.  We reduced the ACIS-S data with the {\tt CIAO} software package (v4.12) and relative calibration files (CALDB 4.9.3), applying standard ACIS data filtering \citep{fruscione2006}. 

We refined the CXO absolute astrometry by cross-matching the  0.5-8 keV X-ray sources blindly detected with \texttt{wavdetect} with optical sources in SDSS DR9. After the re-alignment we find evidence for statistically significant X-ray emission at coordinates RA=$22^{\rm{h}}20^{\rm{m}}02^{\rm{s}}.036\pm0^{\rm{s}}.078$ and $\delta=-02^\circ 50$\arcmin $25$\arcsec$.34\pm 0$\arcsec$.11$, which is consistent with the optical and radio position of AT\,2020xnd. The measured source count-rates and the detection significance are reported in Table \ref{Tab:SoftXray}. At $\delta t_{rest}<25.6\,\rm{d}$ (rest frame) the X-ray source 
shows roughly constant flux. The source experienced significant fading at  $\delta t_{rest}\ge 26\,\rm{d}$, and the flux decays as $F_x\propto t^{-\alpha}$ with $\alpha\approx 4.5$ (Figure \ref{Fig:XrayFBOTS}). This very steep late-time X-ray flux decay is similar to AT\,2018cow (Figure \ref{Fig:XrayFBOTS}), which is the only other FBOT with X-ray observations at these epochs \citep{Margutti19, RiveraSandoval18}. These properties make \sn{} the third FBOT with luminous X-ray emission $L_x > 10^{42}\,\rm{erg\,s^{-1}}$ (Figure \ref{Fig:XrayFBOTS}) and the third with detected X-rays \citep{Margutti19,Coppejans20}.

For each of the first three CXO epochs we extracted a spectrum with \texttt{specextract} using a 1\arcsec\, region around the X-ray source and a source-free region of 33\arcsec\, 
for the background. The neutral hydrogen column density in the direction of the transient is $\rm{NH_{MW}}=4.8\times 10^{20}\,\rm{cm^{-2}}$ \citep{Kalberla05}. We modeled each spectrum with an absorbed power-law model (\texttt{tbabs*ztbabs*pow} within \texttt{Xspec}; \citealt{arnaud1996}).
We found no statistical evidence for spectral evolution. We thus proceeded with a joint spectral fit. The best fitting power-law photon index is $\Gamma=1.40^{+0.33}_{-0.32}$ and we place a 3\,$\sigma$ limit on the intrinsic absorption column of $\rm{NH_{int}}<3\times 10^{22}\,\rm{cm^{-2}}$. The corresponding 0.3--10 keV fluxes and luminosities are reported in Table \ref{Tab:SoftXray}. We found no evidence for statistically significant X-ray emission at $\delta t_{rest}>60\,\rm{d}$ and we place upper limits on the source count-rate of $\lesssim 10^{-4}\,\rm{c\,s^{-1}}$ (0.5--8 keV) assuming Poissonian statistics as appropriate in the regime of low count statistics. This leads to luminosity limits $L_x<(0.3-0.6)\times 10^{42}\,\rm{erg\,s^{-1}}$ using the spectral parameters and model that best fits the earlier observations (Table \ref{Tab:SoftXray}).

\subsubsection{NuSTAR (3-79 keV)}
\label{subsec:nustar}

We observed \sn{} using the Nuclear Spectroscopic Telescope Array (NuSTAR, 3-79 keV) under the joint CXO-NuSTAR program (PI Matthews; program 22500192; IDs 80701407002, 80701407004, 80701407006; Table \ref{Tab:HardXray}). We reduced the data with \texttt{NuSTARDAS} (v1.9.2) and relative calibration files. We centered a source extraction aperture of 1\arcmin\, at the CXO coordinates and we estimated the background using an annulus of inner and outer radii of 1.1\arcmin and 3\arcmin, respectively. Using Poissonian statistics, we found no evidence for significant emission above the background in the source region. The resulting 3--79 keV count-rate, flux and luminosity limits are listed in Table \ref{Tab:HardXray}. We adopt a power-law spectral model with photon index $\Gamma=1.5$ and a counts-to-flux factor of $1.5\times10^{-10}$ for the spectral calibration. We find $L_x<5\times 10^{43}\,\rm{erg\,s^{-1}}$ (3--79 keV) in the time range probed by our observations, which corresponds to $\delta t_{rest}=T_{0}+25$ to $T_{0}+52$d. For comparison, the hard X-ray Compton hump was detected in AT\,2018cow with hard X-ray luminosities $\approx 10^{43}\,\rm{erg\,s^{-1}}$ at $\delta t_{rest}<15\,\rm{d}$  \citep{Margutti19}.

\begin{figure*}
\gridline{\fig{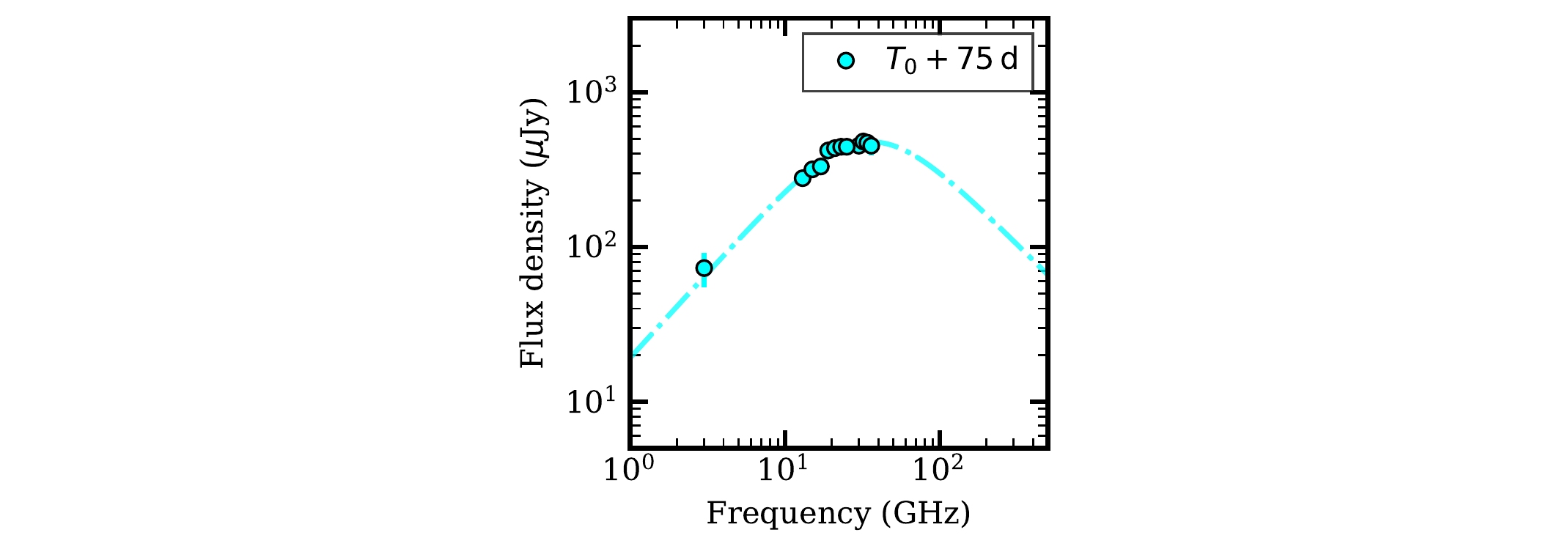}{\textwidth}{(a) Fitting data subset at $\delta t=T_0+75\,\rm{d}$.}
}
\gridline{\fig{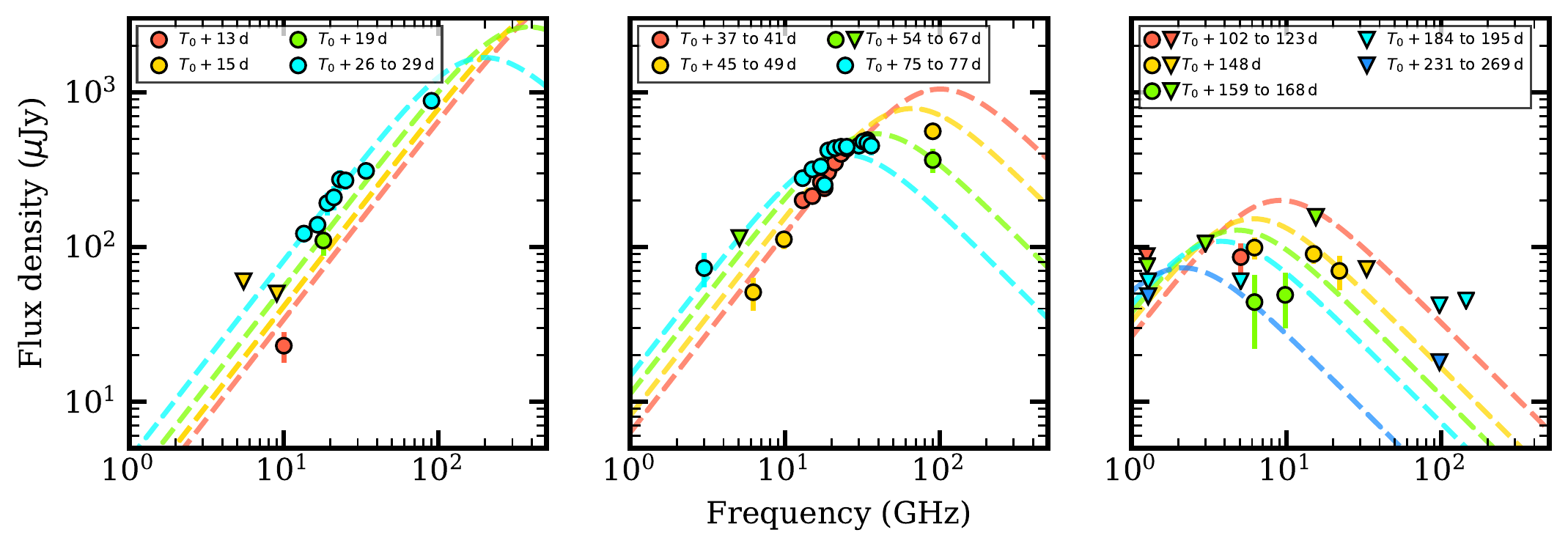}{\textwidth}{(a) Fitting all data.}
}
\gridline{\fig{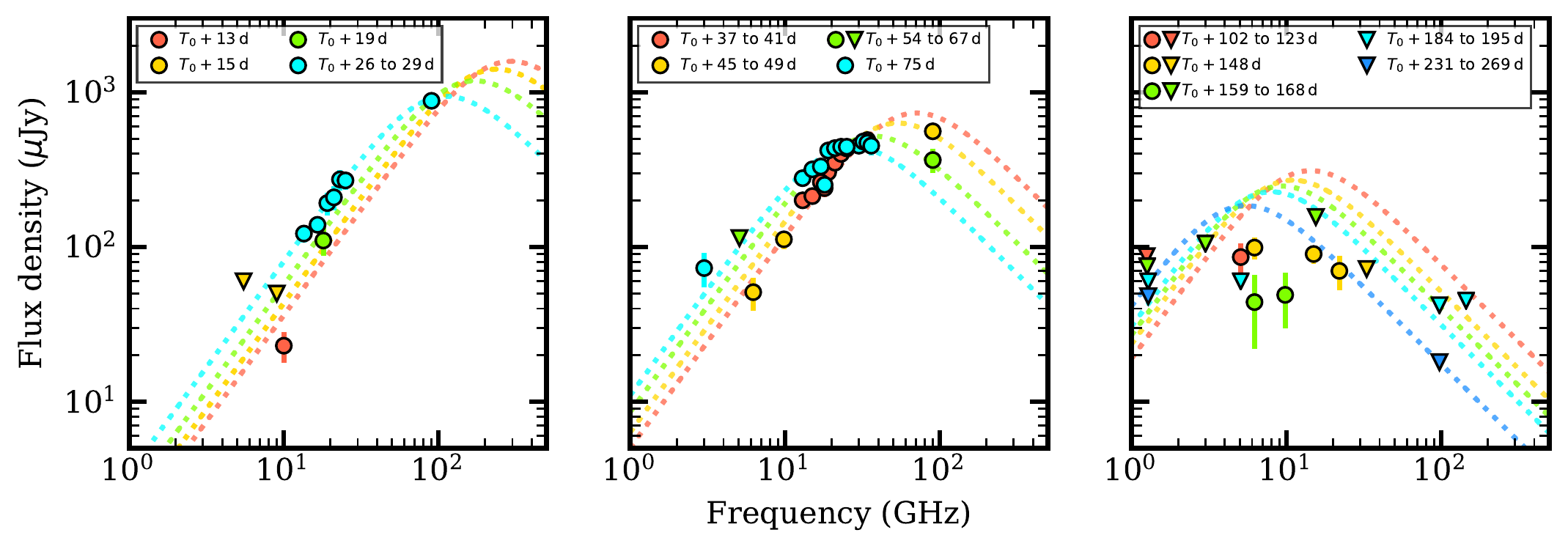}{\textwidth}{(c) Fitting data subset at $\delta t<T_0+75\,\rm{d}$.}
}
\caption{Radio observations of \sn{} taken with the VLA, ATCA, GMRT, GBT, MeerKAT, ALMA, AMI-LA, and eMERLIN radio telescopes, with best fits shown for the model described in §\ref{sec:radio_modelling}. Curves are plotted at the mean time of the observations within the time ranges described in the legend. (a) Our best sampled radio SED, without considering any evolution of the break flux and frequency. (b) All radio observations of \sn{}. The model fit describes the single epoch data well. When considering the entire data set it is clear that the evolving SED model is in marginal contention with the late-time low frequency data from MeerKAT at $\delta t>191$d, over-predicts data at $\delta t=T_{0}+148$d and $\delta t=T_{0}+167$d, while under-predicting the epoch at $\delta t=T_0+75\,\rm{d}$. (c) Radio observations at $\delta t\le T_{0}+75$d. The best fit model parameters and the inferred physical parameters for the data subsets in (a), (b) and (c) are given in Table \ref{Tab:SEDfitRadio} and Table \ref{Tab:SEDfitRadioPhysical}, respectively.\label{fig:radioSED}}
\end{figure*}

\section{Results} \label{sec:results}
\subsection{General Considerations}
\sn{} shows a roughly constant X-ray luminosity at $\delta t_{rest}\le 30\,\rm{d}$ followed by a sharp decline, which was similarly seen in the FBOT AT2018cow (Figure \ref{Fig:XrayFBOTS}) 
Also shown in Figure \ref{Fig:XrayFBOTS} is the X-ray photon index of \sn{}. The measured soft X-ray spectral index $\beta_x= 0.4^{+0.33}_{-0.34}$ (where $F_{\nu}\propto \nu^{-\beta_x}$, \S\ref{SubSec:Chandra}) is shallower than expected from optically thin synchrotron emission above the cooling break frequency $\nu_c$ (\S\ref{sec:synch_cool}). In this regime, $F_{\nu}\propto \nu^{-p/2}$ (e.g., \citealt{Granot02}), where $p$ is the index of the power-law distribution of relativistic electrons with Lorentz factor $\gamma_e$ ($N_e(\gamma_e)\propto \gamma_e^{-p}$). For relativistic shocks of long/short GRBs and the Newtonian shocks of SNe $p\approx 2-3$ (see e.g. \citealt{Chevalier06,fong2019}) hence $F_{\nu}\propto \nu^{-\beta}$ with $\beta=1-1.5$. This again is similar to what was seen in AT2018cow \citep{Margutti19}. While our NuSTAR observations probe the hard X-rays ($3-79$\,\rm{keV}), the large distance of \sn{} only allows us to place  upper limits in this region of the spectrum that are comparable to the luminosity of the observed Compton hump in AT\,2018cow. However, since NuSTAR observations of AT\,2020xnd started at $\delta t_{rest}=25$d, which is after the Compton hump component faded away in AT\,2018cow,  we cannot rule out that \sn{} exhibited a hard X-ray excess similar to the one seen in AT\,2018cow at earlier times \citep{Margutti19}.

Figure \ref{fig:radioSED} shows the evolving radio SED of \sn{} for different subsets of our radio data. At early times ($\delta t\lesssim T_{0}+30$d) the emission is optically thick up to at least $\nu \sim90\,\rm{GHz}$, with a spectral index of $\sim1.4$ below the break. The peak of the SED moves to lower flux density with time, while moving to a lower frequency, until we clearly see the peak in our data at $\delta t\approx T_0+75$d. The optically thin spectral index is not constrained by our data. This spectral shape is similar to the one expected from self-absorbed synchrotron emission, albeit with a flatter self-absorption spectral index (which is expect to be $\sim2$ or $\sim2.5$). The evolution of the break and peak flux suggest the radio emission is from an evolving emitting region that expands and becomes optically thin to lower frequencies with time. This can be interpreted as the result of ejecta material from \sn{} interacting with the CSM surrounding the progenitor. The flatter optically thick slope that we measure could be the result of scintillation effects, however we disfavor this due to the smoothness of the radio SEDs. We are unable to check for short timescale variability or extreme in-band spectral indices (both features of scintillation) due to the low measured flux density of \sn{} at low frequencies.

\subsection{Radio SEDs modeling}\label{sec:radio_modelling}
First, we focus on our radio observations at $\delta t\approx T_0+75$d where the peak of the SED is best sampled. We employ the smoothed broken power-law SED model of Equation \ref{eqn:sed_fit}:
\begin{equation}\label{eqn:sed_fit}
    F(\nu,t)=F_{p}(t)\Bigg[\bigg(\frac{\nu}{\nu_{b}(t)}\bigg)^{-b_{1}/s}+\bigg(\frac{\nu}{\nu_{b}(t)}\bigg)^{-b_{2}/s}\Bigg]^{-s},
\end{equation}
where $\nu_{\rm{b}}$ is the break frequency, $F_{p}$ is the peak flux density at which the asymptotic power-law segments meet, $s>0$ is a smoothing parameter, $b_1$ and $b_2$ are the optically thick and optically thin asymptotic spectral indexes, respectively, at $\nu\ll \nu_{b}$ and $\nu\gg \nu_{b}$. Adopting $s=1$ and $b_{2}=-1$ as typically observed in SNe in the optically thin regime \citep{Chevalier06}, we find $F_{p}=960\pm70\,\mu\rm{Jy}$ and $\nu_{b}=35\pm7\,\rm{GHz}$, respectively, at $T_0+75$d.\footnote{We use the python module \textsc{lmfit} for all fitting performed in this work.} 

Next we attempt to model the entire radio data set shown in panel (b) of Figure \ref{fig:radioSED} assuming a power-law evolution in time of the break frequency ($\nu_{b}(t)=\nu_{b,0}(t/T_{0})^{-\alpha_{\nu}}$) and peak flux density ($F_{p}(t)=F_{p,0}(t/T_{0})^{-\alpha_{F}}$), where our reference epoch is $T_{0}+75\,\rm{d}$. We assume constant spectral indexes $b_1$ and $b_2=-1$, and a common $s=1$. Fitting all of our radio data with in this framework leaves five free parameters: $F_{p,0}$, $\alpha_{F}$, $\nu_{b,0}$, $\alpha_{\nu}$, and  $b_{1}$. As shown in panel (b) of Figure \ref{fig:radioSED} this model is unable to satisfactorily fit the entire set of radio observations. The model under-predicts the peak flux at our best sampled epoch, while over-predicting the later-time data (especially at frequencies close to $10$GHz while being in marginal contention with lower frequency data). This is not unprecedented for FBOTs. AT2018cow also demonstrated a significant change in its radio evolution at $\approx 20$d, with the flux density dropping off markedly faster than predicted by an extrapolation of fits to the early-time data \citep{Margutti19}. We were able to achieve better fitting results when using only the subset of our data at $\delta t \le T_{0}+75$d, with the results shown in panel (c) of Figure \ref{fig:radioSED} and the best fit model parameters given in Table \ref{Tab:SEDfitRadio}. In the following discussion we adopt the model parameters derived for the fits to the subset of data taken at and before $\delta t\sim T_{0}+75\,\rm{d}$ (Figure \ref{fig:radioSED} panel (c)) when inferring physical parameters. This allows us to better compare with our CXO observations (which were taken before $75$d), and account for our best sampled radio SEDs. We will discuss possible interpretations of the late time ($\delta t>T_{0}+75$d) radio flux in \S\ref{sec:radio_late}.

\subsection{Physical Parameters at $\delta t\lesssim T_{0}+75\,\rm{d}$}\label{sec:radio_phys}
Using the results of our fitting in \S\ref{sec:radio_modelling} we can infer physical parameters of the emitting region based on some simple assumptions. Following \citet{Chevalier06}, we calculate the forward shock radius $R_p$, the magnetic field $B_p$, the density of the CSM $n$, and the shock energy $U_p$, all as a function of time and accounting for the significant redshift of \sn{} (which we describe in more details in Appendix \ref{sec:redshift}). We further adopt fiducial values for the shock microphysical parameters of $\epsilon_e=0.1$, $\epsilon_B=0.01$, and an emitting volume fraction $f=0.5$ (compared to a sphere of radius $R_p$).
We let $\alpha\equiv f\epsilon_{e}/(0.5\epsilon_{B})$, which encapsulates the main assumptions of our model.

The radius of the forward shock is 

\begin{subequations}
    \begin{equation}\label{eqn:radius}
    \begin{split}
        R_{p}=4&\times10^{14}\alpha^{-\frac{1}{19}}\bigg(\frac{F_{p}(1+z)^{-1}}{\rm{mJy}}\bigg)^{\frac{9}{19}}\bigg(\frac{D_{\theta}}{\rm{Mpc}}\bigg)^{\frac{18}{19}}\\\times&\bigg(\frac{\nu_{b}(1+z)}{5\,\rm{GHz}}\bigg)^{-1}\,\rm{cm},
    \end{split}
    \end{equation}
    
while its temporal evolution is modeled as: 
    \begin{equation}\label{eqn:radius_scaling}
        R_{p}=R_{p,0}\bigg(\frac{t}{T_{0}}\bigg)^{\alpha_{\nu}-(9/19)\alpha_{F}}=R_{p,0}\bigg(\frac{t}{T_{0}}\bigg)^{1.0\pm0.1}.
    \end{equation}
\end{subequations}

Here $R_{p,0}=(2.8\pm0.2)\times10^{16}$cm is the value of $R_{p}$ at our reference time (which we take to me $\delta t=T_{0}+75$d post explosion). We give statistical errors on the derived scalings, as well as the physical parameters, based on propagating the errors derived from our model fitting. We note, however, that systematic errors resulting from the model assumptions likely dominate. We can then infer the average velocity implied by our model as $(\beta\Gamma c)_{p}=(R_{p}(1+z)/t_{0})$ (which is $0.18\pm0.02$ at $75\,\rm{d}$ post explosion). The best fit model to these data is consistent with no acceleration ($R_{p}\propto t^{1.0\pm0.1}$), although we only confidently measure the location of the peak at the reference epoch. As a comparison, fitting only the data at $\delta t=T_0+75$d, we infer a shock radius of $(2.0\pm0.7)\times10^{16}\,\rm{cm}$ at this epoch, implying an average outflow velocity of the fastest ejecta of $\Gamma\beta c=R(1+z)/t=0.13\pm0.04$. Note that the dependency on the shock microphysical parameters is minor, with an order of magnitude change in $\alpha$ resulting in a $\sim10\%$ change in the radius and velocity (see Equation \ref{eqn:radius}). 

\begin{figure*} 
	\centering
	\includegraphics[width=0.8\textwidth]{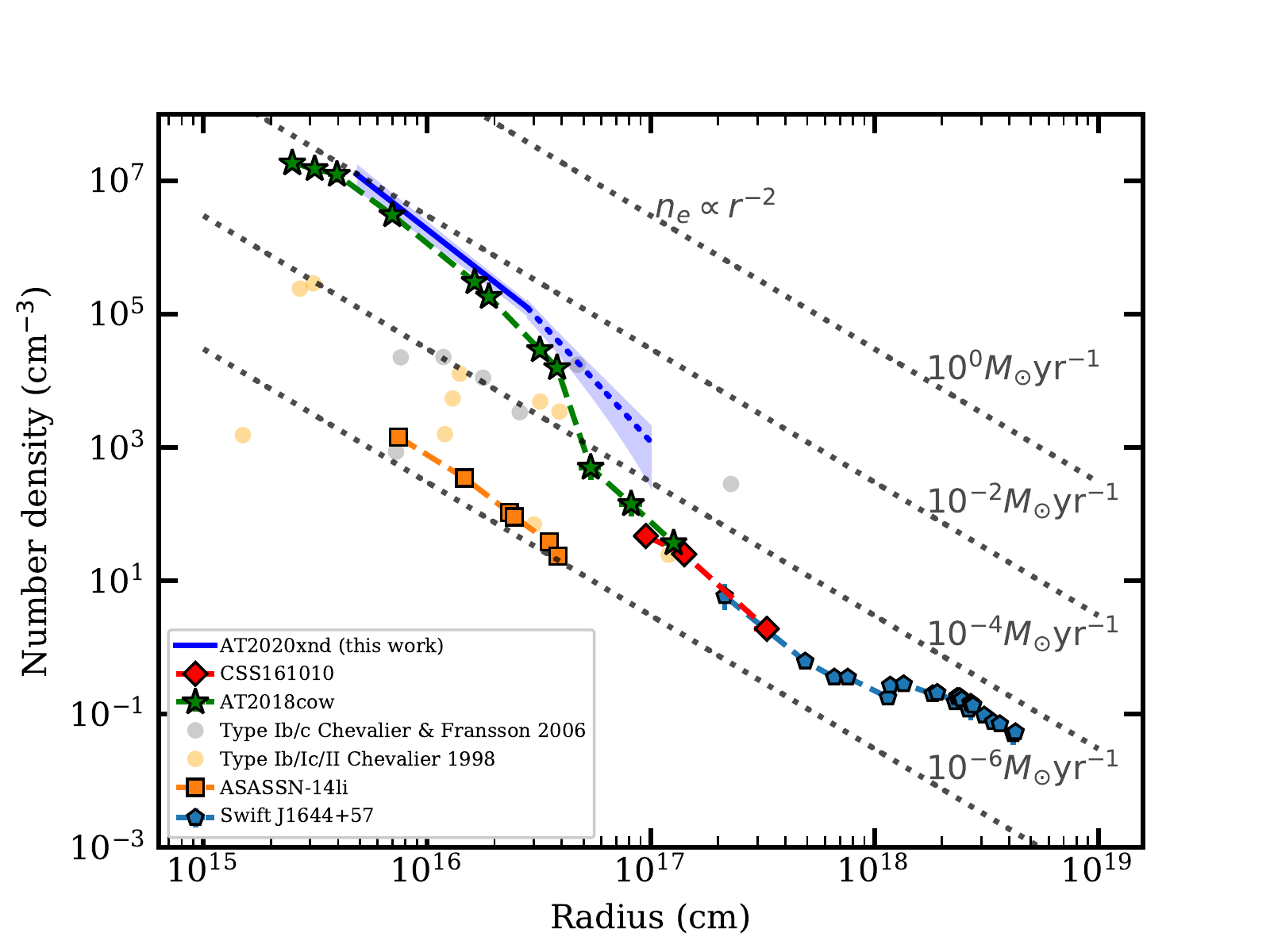}
    \caption{The particle number density as a function of radius for the FBOTs AT2020xnd including the possible steepening at $\delta t\sim T_{0}+75$d (this work), CSS161010 \citep{Coppejans20} and AT2018cow \citep{Margutti19,nayana2018,Ho19}. Also shown are single measurements from various Type Ib/c and Type II SNe where we derived the number density using values from \citet{Chevalier06} and \citet{Chevalier98}. Additionally, we include number density measurements from the prototypical thermal and non-thermal TDEs, ASASSN-14li and SwiftJ1644+57, respectively \citep{Alexander16,Eftekhari18}. Note that the densities derived for the two TDEs used the results of \citealt{BarniolDuran13} as opposed to \citealt{Chevalier06} in this work.  Dotted gray lines show different rates of constant mass loss ($n_{e}\propto r^{-2}$ and assuming a wind velocity of $1000$km\,s$^{-1}$). FBOTs show preferentially high densities and therefore larger mass loss rates compared to the type-Ibc/Ic/II events presented in \citet{Chevalier98} and \citet{Chevalier06}.  
    }
    \label{Fig:number_density}
\end{figure*}

The magnetic field and its scaling are given in Equation \ref{eqn:magnetic_field} and Equation \ref{eqn:magnetic_field_scaling}, respectively:
\begin{subequations}
    \begin{equation}\label{eqn:magnetic_field}
    \begin{split}
        B_{p} = 1.1&\alpha^{-\frac{4}{19}}\bigg(\frac{F_{p}(1+z)^{-1}}{\rm{mJy}}\bigg)^{-\frac{2}{19}}\bigg(\frac{D_{\theta}}{\rm{Mpc}}\bigg)^{-\frac{4}{19}}\\\times&\bigg(\frac{\nu_{b}(1+z)}{5\,\rm{GHz}}\bigg)\,\rm{G}
    \end{split}
    \end{equation}
    
    \begin{equation}\label{eqn:magnetic_field_scaling}
        B_{p}=B_{p,0}\bigg(\frac{t}{T_{0}}\bigg)^{(2/19)\alpha_{F}-\alpha_{\nu}}=B_{p,0}\bigg(\frac{t}{T_{0}}\bigg)^{-1.3\pm0.1}
    \end{equation}
\end{subequations}
Here $B_{p,0}$ is the value of $B_{p}$, the magnetic field associated with the self-absorbed synchrotron emission, at our references time. Note again that, similar to the radius, the value of the magnetic field is relatively robust to the choice of $\alpha$. We find $B_{p,0}=1.04\pm0.07$G. Fitting the SED at $\delta t=T_{0}+75\,\rm{d}$ in isolation we find a post-shock magnetic field of $1.5\pm0.3\,\rm{G}$. These values are similar to those of SNe and other FBOTs (e.g.,  \citealt{Chevalier06,Ho20,Margutti19,Coppejans20}).

Additionally, the peak flux and break frequency allow us to constrain a density profile for the CSM, which was crafted by the mass loss history of the progenitor star, that the SN outflow is interacting with. The density profile of the CSM can be written as $\rho=\dot M/(4\pi r^{2}v_{w})=Ar^{-2}$, where $\dot M$ is the mass-loss rate of the star with wind of velocity $v_{w}$. We define $A_{\star}\equiv A/(5\times10^{11}\,\rm{g}\,\rm{cm}^{-1})$ so that $A_{\star}=1$ for $\dot M=10^{-5}M_{\odot}\rm{yr}^{-1}$ and $v_w=1000\,\rm{km}s^{-1}$, which are typical of Wolf-Rayet stars. Under the assumption that a fraction $\epsilon_{B}$ of the shock energy density is converted to magnetic energy density, $A_{\star}$ can be defined as in Equation \ref{eqn:Astar} and the scaling resulting from our models as in Equation \ref{eqn:Astar_scaling}. 
\begin{subequations}
    \begin{equation}\label{eqn:Astar}
    \begin{split}
        A_{\star p} = &\alpha^{-\frac{8}{19}}\bigg(\frac{\epsilon_{B}}{0.1}\bigg)^{-1}\bigg(\frac{F_{p}(1+z)^{-1}}{\rm{mJy}}\bigg)^{-\frac{4}{19}}\bigg(\frac{D_{\theta}}{\rm{Mpc}}\bigg)^{-\frac{8}{19}}\\&\times\bigg(\frac{\nu_{b}(1+z)}{5\,\rm{GHz}}\bigg)^{2}\bigg(\frac{t(1+z)^{-1}}{10\,\rm{d}}\bigg)^{2}
    \end{split}
    \end{equation}
    
    \begin{equation}\label{eqn:Astar_scaling}
        A_{\star p}=A_{\star p,0}\bigg(\frac{t}{t_{0}}\bigg)^{(4/19)\alpha_{F}-2\alpha_{\nu}+2}=A_{\star p,0}\bigg(\frac{t}{t_{0}}\bigg)^{-0.5\pm0.2}
    \end{equation}
\end{subequations}
For the best fit parameters of our model we find that $A_{\star p,0}=(90\pm10)\times\alpha^{-8/19}(\epsilon_{B}/0.1)^{-1}$ or $A=(1.8\pm0.2)\,\times10^{13}\,\rm{g}\,\rm{cm}^{-1}$ for $\alpha=10$. Using the scaling derived in Equation \ref{eqn:radius_scaling} we see that, under the assumption that the CSM is dominated by fully ionized hydrogen, the number density profile is:
\begin{equation}
    n\bigg(\frac{R_{p}}{R_{p,0}}\bigg)=5\times10^{11}m_{p}^{-1}\bigg(\frac{A_{\star p,0}}{R_{p,0}^{2}}\bigg)\bigg(\frac{R_{p}}{R_{p,0}}\bigg)^{\frac{22\alpha_{F}-76\alpha_{\nu}+38}{19\alpha_{\nu}-9\alpha_{F}}}\,\rm{cm}^{-3},
\end{equation}
and therefore $n=n_{p,0}(R_{p}/R_{p,0})^{-2.5\pm0.2}$. From the previously calculated values of $A_{\star p,0}$ and $R_{p,0}$, we see that $n_{p,0}=(1.2\pm0.2)\times10^{4}\,\rm{cm}^{-3}$ at a radius of $R_{p}\sim2.8\times10^{16}\,\rm{cm}$ (i.e. $R_{p,0}$). These values imply effective mass-loss rates $\dot M\sim3\times 10^{-3}\,M_{\odot}\rm{yr}^{-1}$ for $v_w=1000\,\rm{kms}^{-1}$.  We compare the density profile of \sn{} with other classes of SNe and FBOTs in Figure \ref{Fig:number_density}. The inferred density of the environment of AT2020xnd is very similar to that of the FBOT AT2018cow, and denser than typical SN environments. While similarly large densities can be found around supermassive black holes (SMBHs), dormant intermediate-mass BHs are not expected to be surrounded by high-density media. This provides challenges for FBOT models invoking a TDE around an IMBH (e.g., \citealt{Kuin19,Perley19}). By the end of our observing campaign we are probing emission at $\sim10^{17}$cm. Taking our number density profile and integrating it out to this radius implies a total mass swept up of $\sim10^{-2}\rm\Msol$.

Finally, we calculate the shock energy according to 

\begin{subequations}
    \begin{equation}
    \begin{split}
        U_p=&\bigg(\frac{fR^{3}B^{2}}{6\epsilon_{B}}\bigg)=1.3\times10^{43}f\epsilon_{B}^{-1}\alpha^{-\frac{11}{19}}\bigg(\frac{F_{p}(1+z)^{-1}}{\rm{mJy}}\bigg)^{\frac{23}{19}}\\&\times\bigg(\frac{D_{\theta}}{\rm{Mpc}}\bigg)^{\frac{46}{19}}\bigg(\frac{\nu_{p}(1+z)}{5\,\rm{GHz}}\bigg)^{-1}\,\rm{erg}
    \end{split}
    \end{equation}
    
    \begin{equation}
        U_p=U_{p,0}\bigg(\frac{t}{t_{0}}\bigg)^{\alpha_{\nu}-(23/19)\alpha_{F}}=U_{p,0}\bigg(\frac{t}{t_{0}}\bigg)^{0.5\pm0.3}
    \end{equation}
\end{subequations}

and find that, $U_p=(1.5\pm0.1)\times10^{49}f\epsilon_{B}^{-1}\alpha^{-11/19}\,\rm{erg}$ or  $U_p=(2.0\pm0.2)\times10^{50}$erg, for our fiducial $f$, $\alpha$ and $\epsilon_B$ parameters.  For comparison, the single 75d-SED fit returned a total shock energy of $U=(1.5\pm0.4)\times10^{50}\,\rm{erg}$. For comparisons, AT2018cow showed significantly less energy ($U\sim5\times10^{47}$erg) at similar epochs (\citealt{Margutti19}, Coppejans et al., in prep.), while the mildly relativistic FBOT CSS161010 was powered by a similarly energetic outflow with $U=(2.9\pm0.1)\times10^{49}$erg \citep{Coppejans20}.

\begin{figure*}
	\centering
	\includegraphics[width=0.7\textwidth]{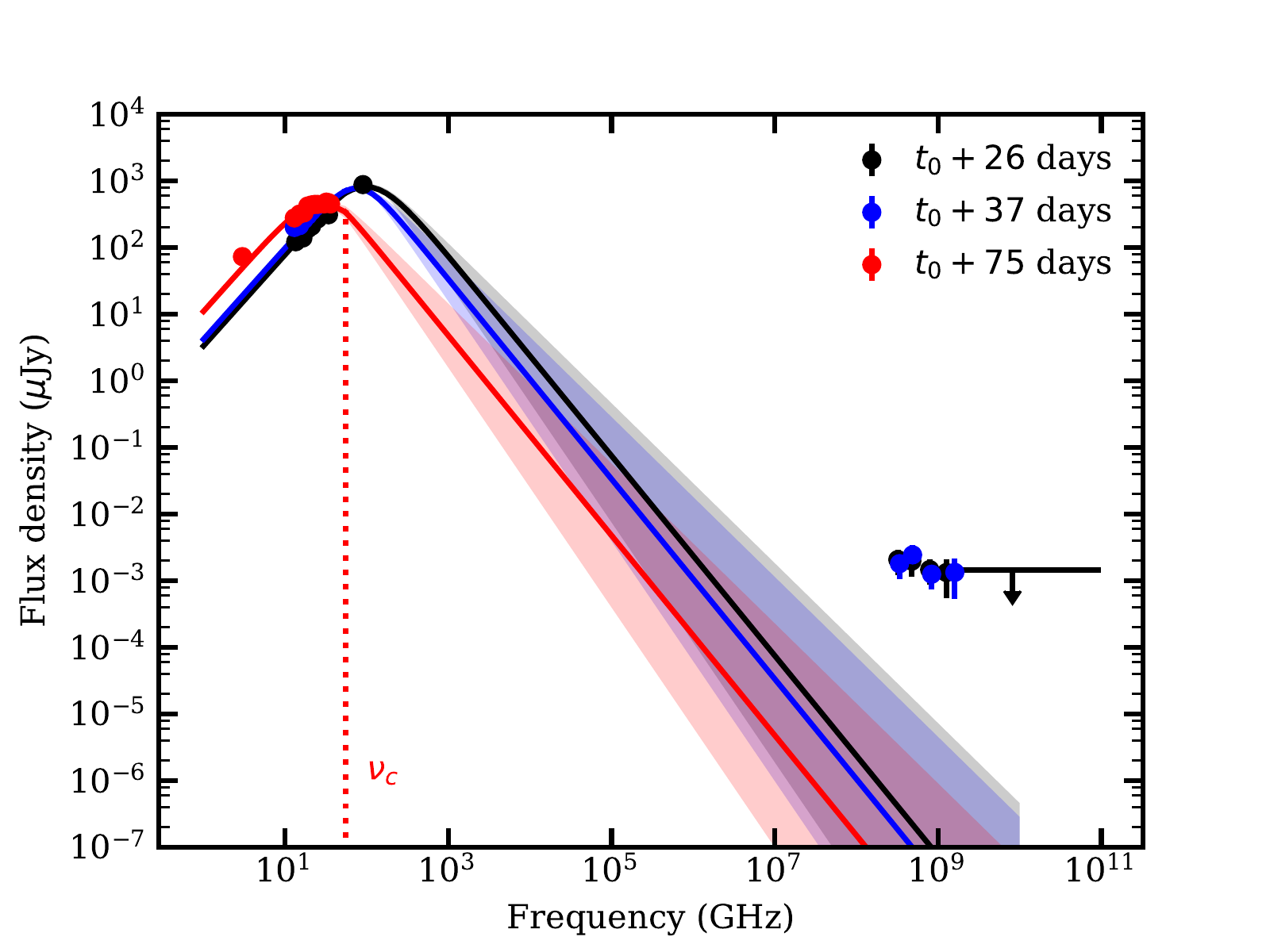}
    \caption{The radio through X-ray SED of \sn{} showing our two \emph{Chandra} epochs and the two nearest-in-time VLA epochs. Also shown is the reference epoch where we best sample the peak of the radio SED. Under the assumption that the peak of the SED is the result of synchrotron self-absorption we estimate the location of the synchrotron cooling break, $\nu_{c}$, at the reference epoch, and demonstrate this here. Due to the magnetic field scaling derived in Section \ref{sec:synch_cool} the cooling break should move to higher frequencies with time. At the black and blue epochs we are in the regime where $\nu_{m},\nu_{c}<\nu_{sa}$, and the cooling break is not seen. Instead we show these epochs with the slope post self absorption frequency steepened from $-(p-1)/2$ to $-p/2$. The wide upper limit given at epoch 1 is from \emph{NuSTAR} and includes the entire 3 to $79$keV observing band. Constraints at the other epochs were comparable. While we fix the post-break spectral index to -1 in our model, the shaded regions show the effect of including an error of $\pm0.3$ on this value.}
    \label{Fig:RadioXraySED}
\end{figure*}

\subsection{Late time emission at $\delta t>75\,\rm{d}$: sub-mm and low frequency constraints}
As discussed in the previous sections, a single power-law evolution of the peak flux and peak frequency with time is unable to provide an accurate representation of the entire data set. Our data coverage is sparser at $\delta t>T_{0}+75\,\rm{d}$, however the data indicate a significantly faster evolution. In this time range we find $\alpha_\nu=2.1\pm0.4$ and $\alpha_F=2.2\pm0.2$, which we calculate by fitting just the SEDs at $\delta t=T_{0}+75$d and $\delta t=T_{0}+148$d. Interestingly, these imply a steepening of the density profile as $n\propto (R/R_{0})^{3.6\pm0.6}$ around $R_p\sim3\times10^{16}$cm. A similar steepening was inferred for the environment of AT2018cow and might represent a defining characteristics of the class of luminous FBOTs. We will discuss possible origins of this density profiles in \S\ref{sec:radio_late}.

\begin{figure*} 
	\centering
	\includegraphics[width=0.50\textwidth]{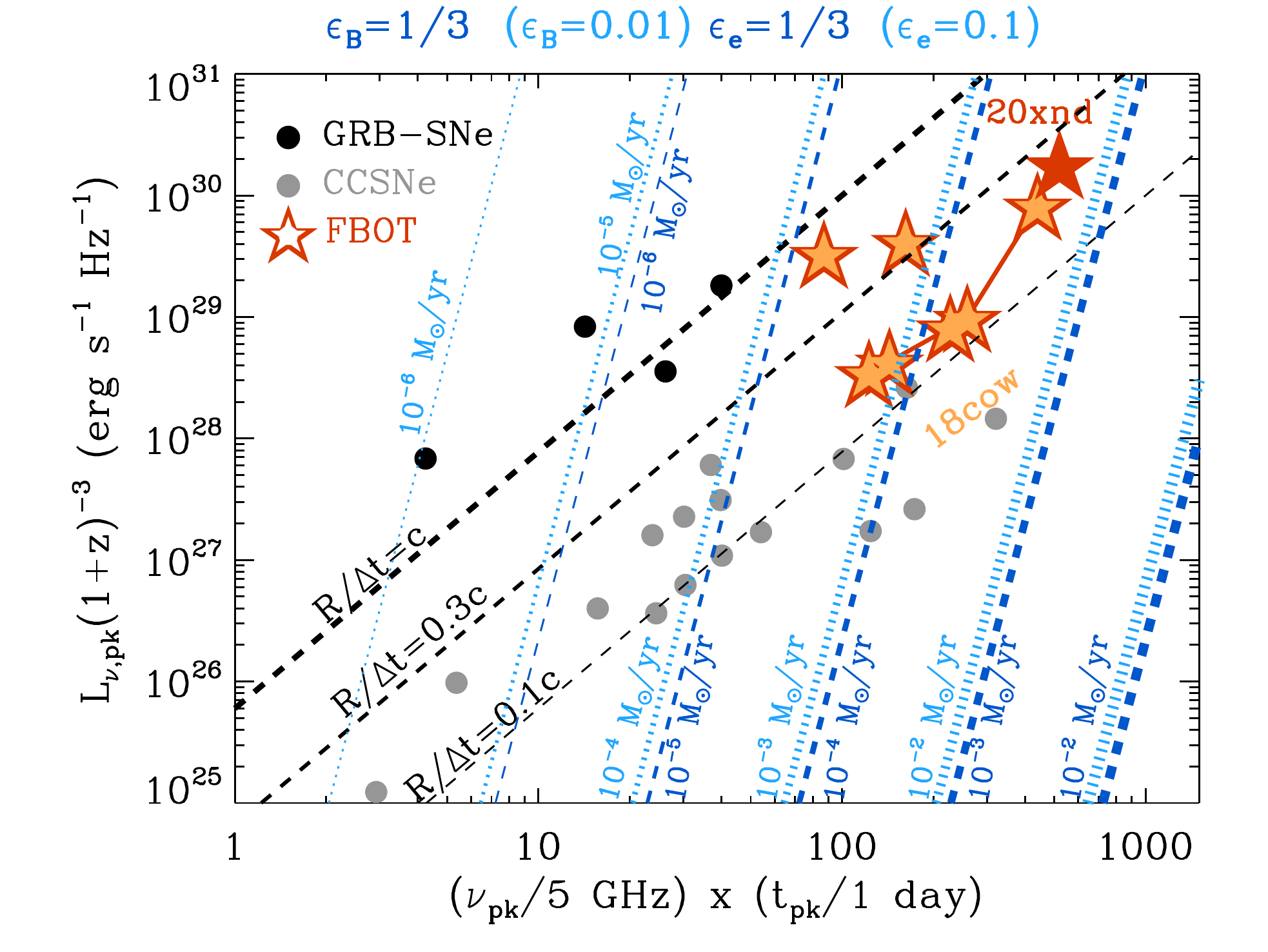}
	\includegraphics[width=0.48\textwidth]{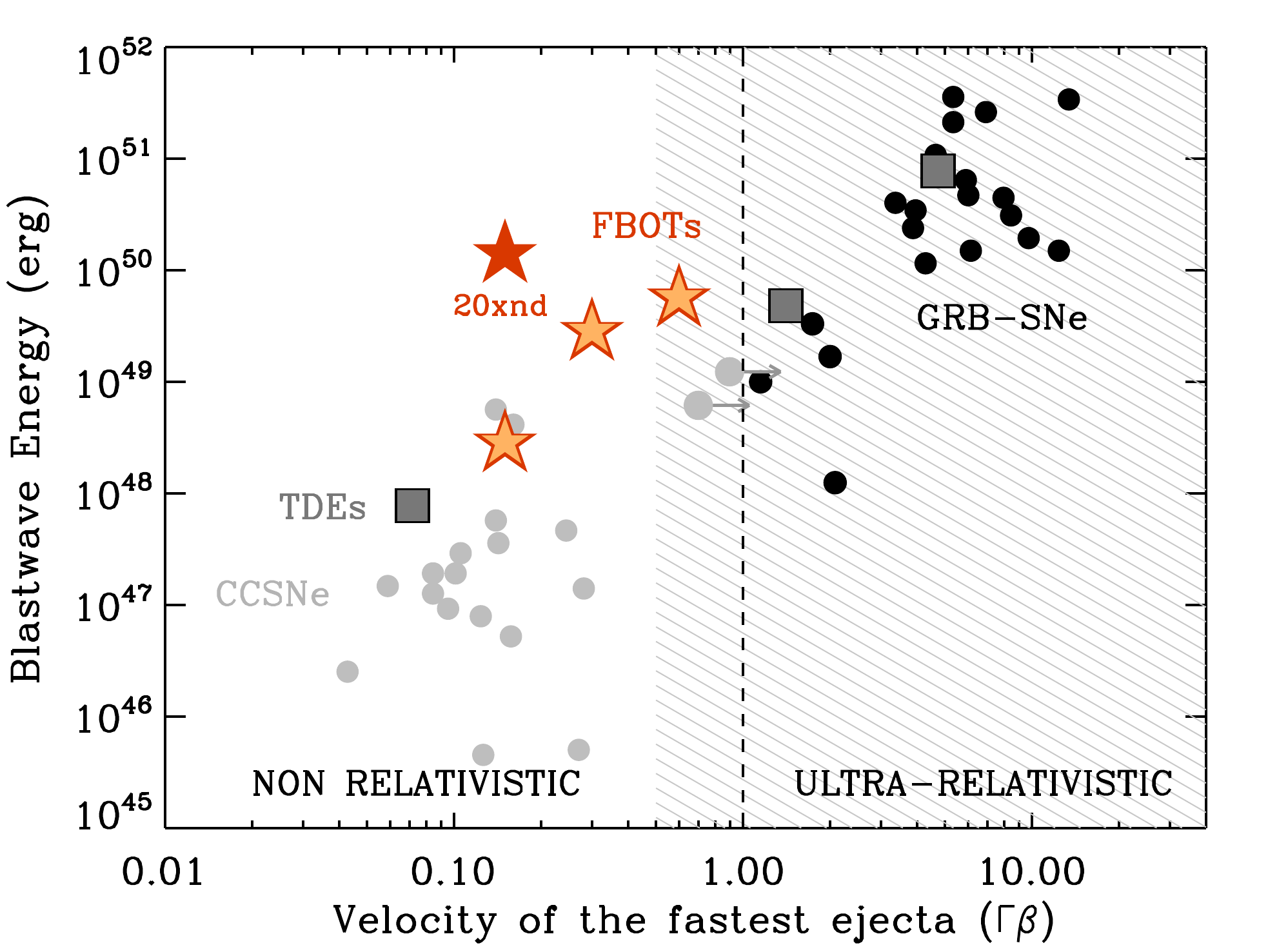}
    \caption{\emph{Left Panel:} Radio phase-space of stellar explosions, where $L_{\rm{\nu,pk}}$ is the spectral peak luminosity, $\nu_{\rm{pk}}$ is the spectral peak frequency and $t_{\rm{pk}}$ is the peak time. Normal core-collapse stellar explosions (CCSNe, gray filled circles) typically show a shock velocity $\sim0.1$c, while  GRB-SNe (Black filled circles) are characterized by significantly faster shocks.  The new class of multiwavelength FBOTs (stars) is currently comprised of four elements and  shows a variety of shock velocities that bridge the gap between  normal SNe (e.g., AT\,2018cow) and the ultra-relativistic jets of GRBs (e.g., CSS\,161010, AT\,2018lug and AT\,2020nxd).   Dashed  dark (dotted light) blue lines: lines of constant equivalent mass-loss rate for an assumed wind velocity $v_w=1000\,\rm{km\,s^{-1}}$ for equipartition parameters $\epsilon_B=\epsilon_e=1/3$ ($\epsilon_B=0.01$, $\epsilon_e=0.1$).
    Black dashed lines: lines of constant apparent velocity $R/ \Delta t$ under the assumption of equipartition.  The factor $(1+z)^{-3}$ accounts for the difference between the angular distance and the luminosity distance. \emph{Right Plot: }  Shock energy vs. fastest ejecta velocity for a variety of transients, including TDEs, SNe and FBOTs. FBOTs represent a new class of transients with energetic shocks that can be transrelativistic. Among FBOTs, AT2020xnd shows the largest equipartition energy. Same color coding as panel on the left. Shaded region: parameter space of engine driven explosions. References: \cite{Margutti19, Ho19, Coppejans20}.
    }
    \label{Fig:RadioPhaseSpace}
\end{figure*}

\subsection{Synchrotron Cooling}\label{sec:synch_cool}
The frequency of the cooling break (where the electrons are radiating a significant fraction of their energy to synchrotron radiation on a dynamical timescale) can be estimated as
$\nu_{c}=(18\pi m_{e}ce)/(t^{2}\sigma_{t}^{2}B^{3})$ where we can use our estimates for the magnetic field made in the previous section. For our best fit model parameters (for data taken at $\delta t\leq T_{0}+75\,\rm{d}$) we see that $\nu_{c}=\nu_{c,0}(t/t_{0})^{1.6\pm0.4}\,\rm{GHz}$, with $\nu_{c,0}=(70\pm10)\,\rm{GHz}$ (we stress however that these calculations are only strictly valid for a constant magnetic field and depend quite strongly on $\alpha$). At the same time the synchrotron frequency is $\nu_m \equiv  \gamma_m^2 \big( \frac{eB}{2\pi m_e c} \big)\approx3\gamma_{m}^{2}$MHz, so we have that $\nu_m<\nu_{sa}<\nu_{c}$ for $\gamma_{m}\lesssim170$. The synchrotron spectrum from a population of electrons accelerated into a power-law distribution with an index $p$ governing their energy distribution steepens from $F_{\nu}\propto \nu^{-(p-1)/2}$ to $F_{\nu}\propto\nu^{-p/2}$ above the cooling break (in the event that $\nu_{sa}<\nu_{m}<\nu_{c}$ or $\nu_{m}<\nu_{sa}<\nu_{c}$, see e.g. \citealt{Granot02}). 

We show the cooling break in the context of the radio through X-ray SED in Figure \ref{Fig:RadioXraySED} for data taken at our reference epoch, along with the two epochs with quasi-simultaneous CXO  and radio observations. At early times $\nu_c<\nu_{sa}$ 
(which might be contributing to ``broadening'' the synchrotron cooling break) and the optically-thin spectrum never demonstrates the $F_{\nu}\propto \nu^{-(p-1)/2}$ scaling, instead moving directly to the $F_{\nu}\propto\nu^{-p/2}$ regime. Accounting for the presence of synchrotron cooling demonstrates that the X-rays are in a significant excess to the radio data taken at approximately the same time, as found for the other FBOTs with both radio and X-ray data, CSS161010 and AT2018cow (\citealt{Coppejans18,Margutti19}). The presence of a luminous X-ray excess of emission appears to be a defining property of optically-luminous FBOTs (further discussed in \S\ref{SubSec:XExcess}).

\section{Discussion} \label{sec:discussion}
\subsection{An X-ray Excess}
\label{SubSec:XExcess}
We detected \sn{} on three epochs with the Chandra X-ray Observatory, with a significant drop occurring between the our second and third observations (see Figure \ref{Fig:XrayFBOTS}). We show the radio through X-ray SED of \sn{} in Figure \ref{Fig:RadioXraySED}, demonstrating the location of the cooling break at various epochs. It is clear that, as for the other two FBOTs with both X-ray and and radio detections, the extrapolated radio spectrum predicts less X-ray emission than observed. This is the case even considering relatively large error on the post break spectral index. From our Chandra observations of \sn{} we can measure the photon index to be $\Gamma=\beta+1=1.4^{+0.33}_{-0.33}$. This is shallower (harder) spectral than we would expect for the fast cooling regime in the X-rays ($\Gamma=2$ to $2.5$), further suggesting a different origin for the X-ray emission in addition to  (and consistent with) our assessment based on the X-ray flux level. Similar considerations based on similar observational evidence, led \cite{Margutti19} to suggest the presence of a centrally-located source of hard X-ray emission in AT2018cow. 

In stark contrast to normal SNe (see e.g., SN2014C; \citealt{Margutti17}), the X-ray emission in AT2018cow is of clear \emph{non-thermal} origin \citep{Margutti19}. An intriguing possibility is that the source of energy of AT2018cow might be connected to energy released by a newly-formed compact object, either in the form of a black-hole or neutron star. With $L_x\approx 6\times 10^{42}\,\rm{erg\,s^{-1}}$ (Figure \ref{Fig:XrayFBOTS}),  the measured X-ray luminosity of \sn{} is $\approx$5 orders of magnitude larger than the Eddington luminosity for a stellar mass black hole (see Table \ref{Tab:SoftXray}). This is similar to what was seen in the FBOTs AT2018cow and CSS161010 (\citealt{Margutti19, Coppejans18}), confirming that FBOTs with associated X-ray emission have a similar luminosity to TDEs and GRBs in the local universe, and are significantly more luminous than ``normal'' CCSNe (see e.g., \citealt{Margutti13,Margutti13b,Drout14,Vinko15,Margutti17,Eftekhari18} for luminosities of these various transient classes).

If the X-ray emission from FBOTs is powered by a nebula of relativistic particles energized by a central engine, the sudden drop in the X-ray luminosity could in principle arise from an abrupt shift in the spectral energy distribution of the emission rather than change in the bolometric decay of the engine's luminosity (which is generally predicted to be smooth in fall-back accretion or magnetar-powered scenarios).  For example, in their recent self-consistent Monte Carlo calculations of jet/magnetar-powered nebulae and engine-powered transients, \citet{Vurm&Metzger21} predict that, for FBOT-scaled engine and ejecta properties, an abrupt increase in the mean particle per energy is predicted to occur on a timescale of $\sim 1$ month after the explosion due to a change in the processes regulating the mass-loading of the wind/jet (namely, the cessation of $\gamma\gamma$ pair production).  The resulting sudden increase in the peak of the non-thermal synchrotron or inverse Compton emission could act to shift the radiated energy from the soft X-ray to the gamma-ray band (where it would go undetected at these relatively late epochs).

\subsection{Late Time Radio Emission}\label{sec:radio_late}
Interestingly, as with the other radio-loud FBOTs with well monitored radio emission, we were not able to describe the entire data set of \sn{} with an evolving broken power-law as described by Equation \ref{eqn:sed_fit} (\citealt{Margutti19, Coppejans20}; Coppejans+ in prep.). This is the result of a sharp late-time drop-off in radio flux that we speculate is a  defining characteristic of FBOTs. In normal SNe, $F_p$ tends to remain constant in time, while $\nu_p\propto t^{-1}$, which leads to the characteristic $F_{\nu}\propto t^{-1}$ in the optically thin regime that is appropriate at these late epochs. The markedly different rate of decay of the radio flux density of FBOTs vs. normal SNe after peak can be immediately appreciated from Figure \ref{Fig:XrayFBOTS}. From a physical perspective, this fast evolution can be the result of a steeply decaying environment density at larger radii, moving from $\rho\propto (R/R_{0})^{2.5\pm0.2}$ at $\delta t\lesssim T_{0}+75$d to $\rho\propto (R/R_{0})^{3.6\pm0.6}$ afterwards (determined by fitting the evolution between the SEDs at $\delta t=T_{0}+75$d and $\delta t=T_{0}+148$d (see Figure \ref{Fig:number_density}). This is inferred from the peak flux density decaying faster at later times ($\alpha_F=0.7\pm0.1$ to $\alpha_F=2.2\pm0.2$), and the break frequency moving faster to lower frequencies ($\alpha_{\nu}=1.3\pm0.1$ to $\alpha_{\nu}=2.1\pm0.4$). However, we emphasize that in the context of our synchrotron formalism where the shock microphysics parameters are constant with time, an equivalent interpretation of the steeply decaying $F_p(t)$ and $\nu_p(t)$ parameters would be that of a rapidly evolving $B(R)$. The fast evolution could also be the result of a decelerating blastwave, however the large errors when jointly fitting the $\delta t=T_{0}+75$d and $\delta t=T_{0}+148$d SEDs prevent us from determining the radius evolution between these epochs.

\subsection{Free-Free Absorption}
Due to the high environment densities we derive based on our radio SED modelling, we consider the potential effects of free-free absorption on the observed radio SED which would manifest as a sharp drop-off at low frequencies (and hence and optically thick flux density $F_{\nu}\propto \nu^{\beta}$ with $\beta>5/2$; e.g., \citealt{weiler2002}). Our SED do not show any evidence for free-free absorption. Following \citealt{Margutti19} we have that the optical depth to free-free absorption is

\begin{equation} \label{Eq:taufreefree}
\begin{split}
\tau_{\rm{ff}}\approx&\frac{\alpha_{\rm{ff}}r}{3}\approx10\bigg(\frac{\nu}{10\,\rm{GHz}}\bigg)^{-2}\bigg(\frac{T_{g}}{10^{4}\,\rm{K}}\bigg)^{-3/2}\bigg(\frac{\rm{A}}{100\rm{A_{\star}}}\bigg)^{2}\\&\times\bigg(\frac{v}{0.1c}\bigg)^{-3}\bigg(\frac{t}{\rm{wk}}\bigg)^{-3}.
\end{split}
\end{equation}
Here $\alpha_{\rm{ff}}$ is the absorption coefficient of free-free absorption, $T_{g}$ is the temperature of the absorbing gas, and $v$ is the shock velocity. The subset of our data at $t\lesssim T_{0}+76\,\rm{d}$ shows a spectral index broadly consistent with synchrotron self-absorbed emission (although with a shallower self-absorption  index than the expected $F_{\nu}\propto\nu^{2}$ for $\nu_{m}<\nu<\nu_{sa}$). Free-free absorption will become prevalent when $\tau_{\rm{ff}}\gtrsim1$, so the lack of any obvious spectral drop-off at low frequencies in any of our epochs implies that $\nu_{\rm{ff,\tau=1}}\ll\nu_{p}$. Using the epoch at $t=T_{0}+75\,$d this implies $\nu_{\rm{ff,\tau=1}}\ll40\,\rm{GHz}$.

We can then use this observational constraint and Equation \ref{Eq:taufreefree} to  
place a constraint on the environment density as 

\begin{equation}
    \frac{A}{A_{\star}}\lesssim300\bigg(\frac{T_{g}}{10^{4}}\bigg)^{3/4}v_{0.1}^{3/2}
\end{equation}
at $t=T_{0}+75\,$d. This is consistent with the values derived in \S\ref{sec:radio_phys} for our derived outflow velocities. 

\subsection{Rapid X-ray decline}
We show our Chandra X-ray observations of \sn{} in Figure \ref{Fig:XrayFBOTS} along with those from CSS161010 and the well sampled AT2018cow. A striking feature of the X-ray emission is the significant increase in the rate of flux decay seen at $\sim20\,\rm{d}$ post discovery (in the target's rest frame). AT2018cow initially declined according to $F_{x}\propto t^{-1}$ before a steepening occurred changing the decay rate to $F_{x}\propto t^{-4.5}$. The optical properties of AT2018cow appeared to morph on a similar epoch, with the previously featureless optical spectrum showing H and He emission lines with a corresponding velocity of $\sim4000\,\rm{kms}^{-1}$, and a high photosphere temperature that showed little evolution (\citealt{Perley2021}). Based on the optical data presented by \cite{Perley2021}, \sn{} appears to show similar properties, although a lack of optical spectra taken around $20\,\rm{d}$ post discovery prevents us from confirming the link between X-ray and optical evolution.

The X-ray fading seen in AT2018cow was in no way smooth, with a high level of variability seen on sub-day timescales (\citealt{Margutti19}), however due to the relative faintness of \sn{} we could not perform a similar analysis in this work.
Sudden changes in the X-ray properties of engine-powered transients/variables are commonly seen in X-ray binary (XRB) systems as sources transition between accretion states \citep{fender2004, remillard2006}. XRBs, however, have X-ray luminosities (attributed to accretion) at approximately, or considerably less than, the Eddington limit, with changes in the X-ray hardness/luminosity (as well as their radio properties, see e.g., \citealt{bright2020}) seen at $\sim1\%\,\rm{L}_{\rm{Edd}}$. Due to the extremely high X-ray luminosity of the FBOTs (e.g. many orders of magnitude above the Eddington limit for \sn{}) attributing the evolving X-ray evolution of AT2018cow/\sn{} to a similar mechanism is challenging as accretion rates this large are simply not seen in Galactic transients, with only the ultraluminous X-ray sources exceeding the Eddington limit.

\section{Conclusions}

\sn{} is now the fourth FBOT to be detected at radio frequencies, and the third at X-rays. Our analysis of the evolving radio and X-ray emission allows us to draw a number of key conclusions on the nature of this object:

\begin{itemize}
    \item The fastest outflows produced by \sn{} (probed through our radio observations) traveled at a significant fraction of the speed of light ($0.1c$ to $0.2c$), a result that is robust to the subset of data which we fit, and also only weakly depends on the model assumptions.
    \item Our observations strengthen the case for FBOTs hosting a central engine. Our X-ray observations are both spectrally harder than, and in excess of, extrapolations of our fits to the radio data. This implies a distinct emission component producing X-rays in \sn{}. Despite observing \sn{} in the hard X-ray with \emph{NuSTAR} we were unable to place meaningful constraints on any hard X-ray excess, as seen in AT2018cow.
    \item Similar to AT2018cow, the X-ray emission from \sn{} underwent a marked change in its decay rate at $\sim20\,\rm{d}$ post explosion. While the physical cause of this evolution remains unclear, it motivates further X-ray studies of FBOTs.
    \item We see a distinct change ($\alpha_F=0.7\pm0.1$ to $\alpha_F=2.2\pm0.2$; $\alpha_{\nu}=1.3\pm0.1$ to $\alpha_{\nu}=2.1\pm0.4$) in the late-time evolution of the radio SED from \sn{}, revealed through our inability to find a satisfactory fit to the entirety of our radio observations with a single evolving synchrotron spectrum. Our attempts led to a model under-predicting our most well-sampled radio epoch, and moderately over-predicting the late-time data at low frequencies. A similar phenomena was seen (and more clearly) in AT2018cow (Coppejans+ in prep.).
    \item We find broad agreement with the results obtained by Ho et al.+2021 for their independent analysis of a separate data set from \sn{}. This includes a similar shock velocity, energy, and steep density profile, with a change in shock properties occurring at around $\delta t=T_{0}+75$d. This change in parameters also prevented them from fitting a single evolving synchrotron self-absorbed SED. Furthermore, they also find X-ray emission in excess of an extrapolation of the radio emission, thus requiring a separate emission component.
\end{itemize}

These properties continue to solidify FBOTs as a new and distinct class of extragalactic transient with luminous counterparts outside the optical spectrum. As surveys such as ZTF and YSE continue to probe events evolving on short timescales only through extensive multi-wavelength followup will the intrinsic nature of fast blue optical transients be revealed.

%

\vspace{5mm}
\facilities{ALMA, ATCA, Chandra, eMERLIN, GBT, GMRT, MeerKAT, NuSTAR, \textit{Swift}, VLA}


\software{astropy, CASA \citep{McMullin07}, MIRIAD \citep{sault1995}, DDFacet \citep{tasse2018}, killMS, matplotlib, numpy, oxkat \citep{heywood_oxkat}, pandas \citep{mckinney-proc-scipy-2010}, WSClean \citep{offringa2014, offringa-wsclean-2017}}


\section*{Acknowledgments}
We thank Anna Ho for sharing an advanced copy of her manuscript with us. Raffaella Margutti's team at Berkeley and Northwestern is partially funded by the Heising-Simons Foundation under grant \# 2018-0911 (PI: Margutti). Support for this work was provided by the National Aeronautics and Space Administration through Chandra Award Number GO1-22062X issued by the Chandra X-ray Center, which is operated by the Smithsonian Astrophysical Observatory for and on behalf of the National Aeronautics Space Administration under contract NAS8-03060.  R.M. acknowledges support by the National Science Foundation under Award No. AST-1909796 and AST-1944985. R.M. is a CIFAR Azrieli Global Scholar in the Gravity \& the Extreme Universe Program, 2019 and an Alfred P. Sloan fellow in Physics, 2019. 
W.J-G is supported by the National Science Foundation Graduate Research Fellowship Program under Grant No.~DGE-1842165 and the IDEAS Fellowship Program at Northwestern University. W.J-G acknowledges support through NASA grants in support of {\it Hubble Space Telescope} programs GO-16075 and GO-16500.
The Berger Time-Domain Group at Harvard is supported in part by NSF and NASA grants.
D.~M.\ acknowledges NSF support from grants PHY-1914448 and AST-2037297.

This project makes use of the NuSTARDAS software package. 
We thank Rob Beswick and the eMERLIN team for approving and carrying out DDT observations of \sn{}.
We thank Jamie Stevens and the ATCA team for approving DDT observations of \sn{}. The Australia Telescope Compact Array is part of the Australia Telescope National Facility which is funded by the Australian Government for operation as a National Facility managed by CSIRO. We acknowledge the Gomeroi people as the traditional owners of the Observatory site.
The National Radio Astronomy Observatory is a facility of the National Science Foundation operated under cooperative agreement by Associated Universities, Inc. GMRT is run by the National Centre for Radio Astrophysics of the Tata Institute of Fundamental Research.
The scientific results reported in this article are based in part on observations made by the Chandra X-ray Observatory. This research has made use of software provided by the Chandra X-ray Center (CXC) in the application packages CIAO.
The MeerKAT telescope is operated by the South African Radio Astronomy Observatory, which is a facility of the National Research Foundation, an agency of the Department of Science and Innovation.
Funding for SDSS-III has been provided by the Alfred P. Sloan Foundation, the Participating Institutions, the National Science Foundation, and the U.S. Department of Energy Office of Science. The SDSS-III web site is http://www.sdss3.org/. SDSS-III is managed by the Astrophysical Research Consortium for the Participating Institutions of the SDSS-III Collaboration including the University of Arizona, the Brazilian Participation Group, Brookhaven National Laboratory, Carnegie Mellon University, University of Florida, the French Participation Group, the German Participation Group, Harvard University, the Instituto de Astrofisica de Canarias, the Michigan State/Notre Dame/JINA Participation Group, Johns Hopkins University, Lawrence Berkeley National Laboratory, Max Planck Institute for Astrophysics, Max Planck Institute for Extraterrestrial Physics, New Mexico State University, New York University, Ohio State University, Pennsylvania State University, University of Portsmouth, Princeton University, the Spanish Participation Group, University of Tokyo, University of Utah, Vanderbilt University, University of Virginia, University of Washington, and Yale University.
We thank the staff of the GMRT that made these observations possible. GMRT is run by the National Centre for Radio Astrophysics of the Tata Institute of Fundamental Research.

\appendix

\section{Cosmological Modification to Chevalier (1998)}\label{sec:redshift}
FBOTs occupy a unique region of the distance/outflow-velocity parameter space for radio transients. The inferred velocities are distinctly non/mildly-relativistic, significantly less than those associated with the distant GRBs \citep{BarniolDuran13} or galactic X-ray binaries \citep{fender2019} while having redshifts more comparable to the former and so must be accounted for. The measured flux density (specific-flux) in the observer's frame (non-primed) is related to the specific-luminosity in the source's rest frame (primed) by $F_{\nu}(\nu)=(1+z)L_{\nu}(\nu^{\prime})/(4\pi D_{L}^{2})$ (this is known as a K correction, see e.g. \citealt{meyer2017,condon2018}) where $D_{L}$ is the luminosity distance and $D_{\theta}=D_{L}(1+z)^{-2}$ is the angular diameter distance. We therefore have that $F_{\nu}(\nu)=(1+z)^{-3}L_{\nu}(\nu^{\prime})/(4\pi D_{\theta}^{2})$, where $\nu^{\prime}=\nu(1+z)$ accounts for cosmological redshift. In \cite{Chevalier98} the angular extent of the source is given by $\theta=R/D$, which is the definition of the angular diameter distance. We wish to apply cosmological corrections to the observed quantities we use to make physical inferences on the FS properties, which are the frequency of the spectral break, the time since explosion, and the flux density at the spectral break. The frequency and time are simple, and are given as $\nu^{\prime}=\nu(1+z)$ and $t^{\prime}=t(1+z)^{-1}$ (and therefore $\nu t=\nu^{\prime} t^{\prime}$ is independent of redshift). For the flux density we begin with our earlier definition $F_{\nu}(\nu)=(1+z)L_{\nu}(\nu^{\prime})/(4\pi D_{L}^{2})$, i.e. the flux density we measure corresponds to the luminosity at the redshifted frequency $\nu^{\prime}$ with the factor $(1+z)$ accounting for the compressed bandwidth over which the flux density is measured in the observer's frame. The quantity $F_{\nu}^{\prime}(\nu^{\prime})=L_{\nu}(\nu^{\prime})/(4\pi D_{L}^{2})$ is the flux density of interest for the models of \cite{Chevalier98}, being the flux density measured in the source frame at frequency $\nu^{\prime}$. We therefore have that $F_{\nu}^{\prime}(\nu^{\prime})=F_{\nu}(\nu)(1+z)^{-1}$ where the quantities on the right hand side of the equality are all measurable. Throughout this work we give quantities in the non-primed (observer) frame and provide the appropriate redshift corrections in each formula for clarity.

\section{Observations}

\startlongtable
\begin{deluxetable*}{ccccccc}
\tablecaption{Radio Observations of \sn{}.}
\tablehead{
\colhead{Start Date} & \colhead{Centroid MJD} & \colhead{Phase$^{\rm{a}}$} & \colhead{Frequency} & \colhead{Bandwidth (GHz)} & \colhead{Flux Density$^{\rm{b}}$} & \colhead{Facility}\\
(dd/mm/yy) &  & (d) & (GHz) & (GHz) & ($\mu$Jy) &
}
\startdata
22/10/2020 & 59145.0109 & 13.01 & 10 & 4 & $23\pm5$ & VLA \\  
25/10/20 & 59147.28 & 15.28 & 5.5 & 2 & $<60$ & ATCA \\
25/10/20 & 59147.28 & 15.28 & 9 & 2 & $<50$ & ATCA \\
29/10/20 & 59151.41 & 19.41 & 18 & 4 & $110\pm20$ & ATCA \\
05/11/20 & 59158.22 & 26.22 & 19.09 & 2 & $192\pm32$ & VLA \\
05/11/20 & 59158.22 & 26.22 & 21.07 & 2 & $209\pm37$ & VLA \\
05/11/20 & 59158.22 & 26.22 & 23.05 & 2 & $273\pm41$ & VLA \\
05/11/20 & 59158.22 & 26.22 & 25.03 & 2 & $269\pm35$ & VLA \\
05/11/20 & 59158.74 & 26.74 & 13.49 & 3 & $122\pm15$ & VLA \\
05/11/20 & 59158.74 & 26.74 & 16.51 & 3 & $139\pm16$ & VLA \\
06/11/20 & 59159.82 & 27.82 & 5.07 & 0.512 & $<54$ & eMERLIN \\
07/11/20 & 59160.09 & 28.09 & 90.00$^{\rm{c}}$ & 30 & $900\pm100$ & GBT \\
07/11/20 & 59160.41 & 28.41 & 34 & 4 & $310\pm20$ & ATCA \\
16/11/20 & 59169.11 & 37.11 & 18.98 & 2 & $304\pm31$ & VLA \\
16/11/20 & 59169.11 & 37.11 & 20.99 & 2 & $350\pm31$ & VLA \\
16/11/20 & 59169.11 & 37.11 & 23.01 & 2 & $399\pm32$ & VLA \\
16/11/20 & 59169.11 & 37.11 & 25.02 & 2 & $433\pm33$ & VLA \\
16/11/20 & 59169.14 & 37.14 & 12.97 & 2 & $200\pm23$ & VLA \\
16/11/20 & 59169.14 & 37.14 & 15.00 & 2 & $213\pm17$ & VLA \\
16/11/20 & 59169.14 & 37.14 & 17.02 & 2 & $262\pm23$ & VLA \\
19/11/20 & 59172.27 & 40.27 & 18 & 4 & $240\pm31$ & ATCA \\
24/11/20 & 59177.09 & 45.09 & 9.82 & 4 & $112\pm15$ & VLA \\
24/11/20 & 59177.10 & 45.10 & 6.22 & 4 & $51\pm12$ & VLA \\
25/11/20 & 59178.06 & 46.06 & 90.00$^{\rm{c}}$ & 30 & $560\pm60$ & GBT \\
27/11/20 & 59180.25 & 48.24 & 34 & 4 & $490\pm63$ & ATCA \\
03/12/20 & 59186.05 & 54.05 & 90.00$^{\rm{c}}$ & 30 & $360\pm60$ & GBT \\
15/12/20 & 59198.59 & 66.59 & 5.072 & 0.512 & $<114$ & eMERLIN \\
24/12/20 & 59207.05 & 75.05 & 29.98 & 2 & $451\pm51$ & VLA \\
24/12/20 & 59207.05 & 75.05 & 31.99 & 2 & $480\pm58$ & VLA \\
24/12/20 & 59207.05 & 75.05 & 34.01 & 2 & $471\pm52$ & VLA \\
24/12/20 & 59207.05 & 75.05 & 36.02 & 2 & $451\pm60$ & VLA \\
24/12/20 & 59207.07 & 75.07 & 18.98 & 2 & $420\pm33$ & VLA \\
24/12/20 & 59207.07 & 75.07 & 20.99 & 2 & $435\pm34$ & VLA \\
24/12/20 & 59207.07 & 75.07 & 23.01 & 2 & $444\pm36$ & VLA \\
24/12/20 & 59207.07 & 75.07 & 25.02 & 2 & $444\pm34$ & VLA \\
24/12/20 & 59207.09 & 75.09 & 12.98 & 2 & $278\pm24$ & VLA \\
24/12/20 & 59207.09 & 75.09 & 15.00 & 2 & $317\pm23$ & VLA \\
24/12/20 & 59207.09 & 75.09 & 17.02 & 2 & $331\pm25$ & VLA \\
24/12/20 & 59207.11 & 75.11 & 3 & 2 & $73\pm18$ & VLA \\
26/12/20 & 59209.26 & 77.26 & 18 & 4 & $252\pm36$ & ATCA \\
19/01/21 & 59234.61 & 102.61 & 5.072 & 0.512 & $86\pm19$ & eMERLIN \\
09/02/21 & 59254.36 & 122.36 & 1.25 & 0.4 & $<87$ & GMRT \\
07/03/21 & 59280.69 & 148.69 & 33 & 8 & $<72$ & VLA \\
07/03/21 & 59280.71 & 148.71 & 22 & 8 & $70\pm17$ & VLA \\
07/03/21 & 59280.73 & 148.73 & 15 & 6 & $90\pm12$ & VLA \\
07/03/21 & 59280.75 & 148.75 & 6.224 & 4 & $99\pm16$ & VLA \\
18/03/21 & 59291.44 & 159.44 & 15.5 & 4 & $<156$ & AMI-LA \\
21/03/21 & 59294.33 & 162.33 & 0.75 & 0.4 & $<177$ & GMRT \\
23/03/21 & 59296.33 & 164.33 & 1.26 & 0.4 & $<75$ & GMRT \\
26/03/21 & 59299.77 & 167.77 & 3 & 2 & $<105$ & VLA \\
26/03/21 & 59299.78 & 167.78 & 9.824 & 4 & $49\pm19$ & VLA \\
26/03/21 & 59264.80 & 167.80 & 6.224 & 4 & $44\pm22$ & VLA \\
12/04/21 & 59316.48 & 184.48 & 97.49 & 8 & $<42$ & ALMA \\
12/04/21 & 59316.53 & 184.53 & 144.99 & 8 & $<45$ & ALMA \\
16/03/21 & 59323.94 & 191.94 & 15.5 & 4 & $<135$ & AMI-LA \\
19/04/21 & 59323.18 & 191.18 & 1.28 & 0.856 & $<60$ & MeerKAT \\
23/04/21 & 59327.28 & 195.28 & 5.072 & 0.512 & $<60$ & eMERLIN \\
29/05/21 & 59363.33 & 231.33 & 1.28 & 0.856 & $<48$ & MeerKAT \\
06/07/21 & 59401.31 & 269.31 & 97.44 & 8 & $<18$ & ALMA \\
\enddata
\tablecomments{$^{\rm{a}}$ Days since MJD MJD 59132, using the central time of the exposure on source. $^{\rm{b}}$ Uncertainties are quoted at 1$\sigma$, and upper-limits are quoted at $3\sigma$. The errors take a systematic uncertainty of 5\% (VLA), 15\% (GMRT), 10\% (ATCA), 10\% (eMERLIN), 10\% (GBT), 10\% MeerKAT into account. $^{\rm{c}}$ As MUSTANG-2 is a bolometer instrument sensitive between 75 and $105\,\rm{GHz}$ the central frequency depends on the spectral index of the emission through the band. The impact of this is discussed in further detail in the main text.\label{Tab:radio}}
\end{deluxetable*}

\begin{deluxetable*}{lccc}\label{Tab:radio_cal}
\tablecaption{Calibrators and array configurations used during our radio observations of \sn{}.}
\tablehead{
Instrument & Primary Calibrator(s) & Secondary/Pointing Calibrator(s) & Array Configuration(s)
}
\startdata
ALMA & J2253$+$1608 & J2218$-$0335 & C-5, C-7 \\
AMI-LA & 3C286 & J2226$+$0052 & fixed \\
ATCA & 1934$-$638 & 2216$-$038 & 6B, H168 \\
eMERLIN & 1331$+$3030 & 1407$+$2827, 2218$-$0335 & fixed \\ 
GBT & Neptune & 2225$-$0457 & single dish \\
GMRT & 3C48 & J2212$+$0152 & fixed \\
MeerKAT & J1939$-$6342 & J2225$-$0457 & fixed \\ 
VLA & 3C147, 3C48 & J2218$-$0335, J2225$-$0457 & A, BnA, D \\
\enddata
\end{deluxetable*}

\begin{deluxetable*}{ccccccc}
\tablecaption{\emph{Chandra} X-ray Observations of \sn{}}
\tablehead{
\colhead{Start Date} & \colhead{Phase$^{\rm{a}}$} & \colhead{Exposure} & \colhead{Net Count-Rate} & \colhead{Significance} & \colhead{Flux$^{b}$} & \colhead{Luminosity} \\
& & &(0.5-8 keV)  &  & (0.3-10 keV)  & (0.3-10 keV)\\
(UT) & (days)& (ks) &($10^{-4}\,\rm{c\,s^{-1}}$) & ($\sigma$) & ($10^{-14}\rm{erg\,s^{-1}cm^{-2}}$) &  ($10^{42}\rm{erg\,s^{-1}}$)$^{c}$
}
\startdata
2020-11-04 15:51:28 & 25.9 & 19.82 & $12.2\pm 2.5$  &10.8$^{\rm{d}}$ &$3.32^{+0.73}_{-0.75}$ & $6.41^{+1.44}_{-1.41}$ \\
2020-11-10 17:06:55 & 31.8& 19.82 &$11.6\pm 2.5$  &9.9$^{\rm{d}}$ & $3.26^{+0.81}_{-0.77}$ & $6.29^{+1.45}_{-1.61}$ \\
2020-11-25 22:55:07 &46.6 & 19.82 &$1.95\pm 1.00$  &4.4$^{\rm{e}}$ & $0.55^{+0.28}_{-0.28}$ & $1.05^{+0.55}_{-0.55}$ \\
2020-12-24 02:21:05 &75.0 &19.75 &$<1.52$  &$<3$ & $<0.33^{\rm{f}}$ & $<0.60$ \\
2021-04-12 14:05:14 &184.6 &16.86 &\multirow{2}{*}{\hspace{-5.5pt}\Big\}$\,<1.09^{\rm{g}}$} & \multirow{2}{*}{$<3$} & \multirow{2}{*}{$<0.24^{\rm{f}}$} & \multirow{2}{*}{$<0.43$} \\
2021-04-13 02:16:12 &185.1 &19.82 & & & & \\
2021-06-07 04:27:03 & 240.2 & 39.55 & $<0.76$ & $<3$ & $<0.17$ & $<0.30$ \\
\enddata
\tablecomments{$^{\rm{a}}$ Days since MJD 59132, using the middle time of the exposure. $^{\rm{b}}$ Uncertainties are quoted at 1$\sigma$, and upper-limits are quoted at $3\sigma$. Observed Flux . $^{\rm{c}}$ Corrected for Galactic absorption. $^{\rm{d}}$ Blind-detection significance. $^{\rm{e}}$ Targeted-detection significance. $^{\rm{f}}$ Power-law spectral model with $\Gamma=1.4$ is used to convert upper limits from count rates to fluxes. $^{\rm{g}}$ Exposures are merged for a deeper detection limit.\label{Tab:SoftXray}}
\end{deluxetable*}

\begin{deluxetable*}{cccccc}
\tablecaption{\emph{NuSTAR} X-ray Upper Limits of \sn{}.}
\tablehead{\colhead{Start Date} & \colhead{Phase$^{\rm{a}}$} & \colhead{Exposure$^{b}$} & \colhead{Net Count Rate$^{b}$} & \colhead{Obs Flux$^{b}$} & \colhead{Luminosity$^{b}$} \\
 &  & &(3-79 keV)  &  (3-79 keV)  & (3-79 keV)\\
(UT) & (days)& (ks) &($10^{-4}\,\rm{c\,s^{-1}}$) & ($10^{-14}\rm{erg\,s^{-1}cm^{-2}}$) &  ($10^{42}\rm{erg\,s^{-1}}$)\\
}
\startdata
2020-11-04 05:06:09 & 25.2  & 57.73 & $< 16.6$ & $<$ 26.5 & $< 48.5$ \\
2020-11-10 15:40:03 & 31.7 & 62.6 & $<$ 15.4 & $<$ 24.5 & $<$ 44.9 \\
2020-11-30 22:00:05 & 51.9 & 81.8 & $<$ 13.9 & $<$ 22.2 & $<$ 40.6\\
\enddata
\tablecomments{$^{\rm{a}}$ Days since MJD 59132. $^{\rm{b}}$ Exposure of Modules A plus exposure for Module B. We adopt a power-law spectral model with photon index $\Gamma=1.5$ and a counts-to-flux factor of $1.5\times10^{-10}$ (cgs units) for calculating upper limits. \label{Tab:HardXray}}
\end{deluxetable*}

\begin{deluxetable*}{ccccccccc}
\tablecaption{Best fit parameters for different subsets of our radio data, fit with Equation \ref{eqn:sed_fit}. Values given without errors are fixed when performing the fit.}
\tablehead{
Data Subset & $\log(\nu_{b,0})^{\rm{a}}$ & $\log(2^{s}F_{p,0})^{\rm{a}}$ & $b_{1}$ & $b_{2}$ &
$\alpha_{F}$ & $\alpha_{\nu}$ & s & $\chi^{2}_{\nu}$ \\
& (log(GHz)) & (log($\mu$Jy)) & & & & & &
}
\startdata
$t=t_{0}+75$ & $1.54\pm0.09$ & $3.28\pm0.03$ & $1.1\pm0.2$ & $-1$ & 0 & 0 & $1$ & 0.4 \\
all & $1.34\pm0.02$ & $3.20\pm0.01$ & $1.29\pm0.06$ & $-1$ & $1.40\pm0.08$ & $2.03\pm0.09$ & $1$ & $3.1$ \\
$t\leq t_{0}+75$ & $1.39\pm0.03$ & $3.24\pm0.02$ & $1.37\pm0.07$ & $-1$ & $0.7\pm0.1$ & $1.3\pm0.1$ & $1$ & $2.1$ \\
\enddata
\tablecomments{$^{\rm{a}}$ Quantities with a subscript 0 are defined at $T_{0}+75\,\rm{d}$.\label{Tab:SEDfitRadio}}
\end{deluxetable*}

\begin{deluxetable*}{cccccccc}
\tablecaption{Physical parameters as measured at the reference time, see main text for parameter scaling. Values assume $\epsilon_{e}=0.1$, $\epsilon_{B}=0.01$, and $f=0.5$. For the mass loss rate we assume a wind velocity of $1000\,\rm{km}\,\rm{s}^{-1}$. The errors reported here are the result of propagating the uncertainties in the fit parameters reported in \ref{Tab:SEDfitRadio}, they are likely underestimated due to inherent uncertainties associated with the assumptions made in out model. Due to the large uncertainty in the radius for our model fit to the data taken at $75\,\rm{d}$ post explosion, we do not give a density or mass loss.}
\tablehead{Data Subset & $R_{p,0}$ & $B_{p,0}$ & $(\Gamma\beta c)_{p,0}$ & $U_{p,0}$ & $n_{p,0}$ & $\Dot{M}_{p,0}$ & $\nu_{c,p,0}$ \\
& ($\times10^{16}\,\rm{cm}$) & (G) & & ($\times10^{50}$erg) & ($\times10^{5}\,\rm{cm}^{-3}$) & ($\times10^{-3}\,M_{\odot}\rm{yr}^{-1}$) & (GHz)
}
\startdata
$t=t_{0}+75$ & $2.0\pm0.4$ & $1.5\pm0.3$ & $0.13\pm0.03$ & $1.5\pm0.4$ & & & $20\pm10$ \\
all & $2.9\pm0.2$ & $0.96\pm0.05$ & $0.19\pm0.01$ & $1.9\pm0.1$ & $1.0\pm0.2$ & $2.7\pm0.5$ & $70\pm10$ \\
$t\leq t_{0}+75$ & $2.8\pm0.2$ & $1.04\pm0.07$ & $0.18\pm0.01$ & $2.0\pm0.2$ & $1.2\pm0.2$ & $3.2\pm0.7$ & $50\pm10$\\
\enddata
\tablecomments{$^{\rm{a}}$ Quantities with a subscript 0 are defined at $T_{0}+75\,\rm{d}$.\label{Tab:SEDfitRadioPhysical}}
\end{deluxetable*} 


\bibliography{AT2020xnd,master_sne}{}
\bibliographystyle{aasjournal}



\end{document}